\documentclass[a4paper,10pt,twoside]{article}

\usepackage[utf8x]{inputenc}
\usepackage{amsmath}
\usepackage{amssymb}
\usepackage{amsthm}
\usepackage{graphicx}
\usepackage{cite}
\usepackage{pgfplots}
\usepackage{geometry}
\usepackage[authoryear, round]{natbib}
\usepackage{amsfonts}
\providecommand{\U}[1]{\protect\rule{.1in}{.1in}}
\usepackage{fancyhdr}
\usepackage{array}
\usepackage{float}
\usepackage{multirow}
\usepackage{esint}
\usepackage[labelfont=bf]{caption}
\usepackage{subcaption}
\usepackage{amsthm}
\usepackage{tikz}
\usepackage{pgflibraryarrows}
\usepackage{makecell}
\usepackage{arydshln}
\usetikzlibrary{decorations.markings}
\usepackage{titlesec}
\usepackage{booktabs}
\usepackage{xcolor}
\usepackage{hyperref}

\definecolor{node_color}{gray}{0.925}

\titleformat*{\section}{\normalsize\bfseries}
\titleformat*{\subsection}{\normalsize\itshape}

\floatstyle{ruled}
\newfloat{algorithm}{tbp}{loa}
\providecommand{\algorithmname}{Algorithm}
\floatname{algorithm}{\protect\algorithmname}

\newcommand{\stack}[2]{\array{c}{#1}\\[-0.7ex]{#2}\endarray}
\newcommand{\overbar}[1]{\mkern 2.5mu\overline{\mkern-2.5mu#1\mkern-2.5mu}\mkern 2.5mu}
\newcommand{\scal}{\text{\footnotesize$\mathcal{C}$}}
\newcommand{\sscal}{\text{\tiny$\mathcal{C}$}}

\newtheorem{prop}{Proposition}
\newtheorem{theorem}{Theorem}
\newtheorem{definition}{Definition}

\begin{document}

\title{\Large{\textbf{Structure Learning of Contextual Markov Networks using Marginal Pseudo-likelihood}}\footnote{This is the peer reviewed version of the following article: Pensar et al. Structure Learning of Contextual Markov
Networks using Marginal
Pseudo-likelihood. \emph{Scandinavian Journal of Statistics}, Vol. 44: 455 -- 479, 2017, which has been published in final form at \href{https://doi.org/10.1111/sjos.12260}{https://doi.org/10.1111/sjos.12260}. This article may be used for non-commercial purposes in accordance with Wiley Terms and Conditions for Use of Self-Archived Versions.} }

\author{JOHAN PENSAR\\
\normalsize{\emph{Department of Mathematics and Statistics, \AA bo Akademi University}} \\
HENRIK NYMAN \\
\normalsize{\emph{Department of Mathematics and Statistics, \AA bo Akademi University}} \\
JUKKA CORANDER \\
\normalsize{\emph{ Department of Mathematics and Statistics, University of Helsinki}} \\
\normalsize{\emph{ Department of Biostatistics, University of Oslo}} \\
}

\date{}

\maketitle
\vspace{-0.75cm}
\begin{abstract}
\noindent \textbf{ABSTRACT.} Markov networks are popular models for discrete multivariate systems where the dependence structure of the variables is specified by an undirected graph. To allow for more expressive dependence structures, several generalizations of Markov networks have been proposed. Here we consider the class of contextual Markov networks which takes into account possible context-specific independences among pairs of variables. Structure learning of contextual Markov networks is very challenging due to the extremely large number of possible structures. One of the main challenges has been to design a score, by which a structure can be assessed in terms of model fit related to complexity, without assuming chordality. Here we introduce the marginal pseudo-likelihood as an analytically tractable criterion for general contextual Markov networks. Our criterion is shown to yield a consistent structure estimator. Experiments demonstrate the favorable properties of our method in terms of predictive accuracy of the inferred models.
\\
\\
\noindent {\small \emph{\textbf{Key words:}} Bayesian inference, context-specific independence, graphical models, Markov networks, pseudo-likelihood, structure learning}
\end{abstract}

\section{Introduction}
Markov networks, also known as undirected graphical models, are popular tools for numerous application fields in science and technology \citep{Lauritzen96, Koller09}. In statistical physics, examples of this kind of interaction models for discrete systems are the classical Ising and Potts models for finite lattices \citep{Sherrington75, Ekeberg13}. In social sciences, graphical log-linear models provide the basic approach for scrutinizing dependences among variables in high-dimensional contingency tables \citep{Lauritzen96, Edwards00}. Image analysis and spatial statistics represent other research fields where Markov random fields are widely used to encode neighborhood dependence structures in data \citep{Tjelmeland98}.

Markov networks enjoy particularly tractable properties when the corresponding graph is chordal, since this allows for a full factorization of the joint distribution using maximal clique marginals and the separator sets of the cliques \citep{Lauritzen96}. As a consequence, structural learning of the graph is also greatly facilitated by chordality and a majority of the graph learning methods have therefore been developed under this assumption \citep{Dawid93, Madigan94, Corander08, Corander13}. Nevertheless, chordality would rarely be supported by the theory underlying a particular application, instead it is mostly assumed on the basis of mathematical and computational convenience. 

In computational physics, image analysis, and spatial statistics, non-chordal graphs are a commonplace. Consequently, efficient inference approximations have been developed over the years to resolve the difficulties arising from the intractable partition function of a multivariate distribution for which the dependence structure is defined by a non-chordal graph. Pseudo-likelihood and numerous variational approximations are the most widely applied approximate inference techniques for such models \citep{Besag75, Wainwright08}. In addition to using pseudo-likelihood for inferring the model parameters \citep{Ekeberg13}, it has also been used in several structure learning methods \citep{Csiszar06,Hofling09,Ravikumar10}.

While the Markov properties of undirected graphical models are useful for obtaining a lower-dimensional representation of a multivariate distribution, they may also obscure more local forms of independence. Context-specific independences between pairs of nodes given a particular instantiation of their neighbors will be neglected in a Markov network, and hence, various incarnations of such models have been proposed \citep{Corander03,Hojsgaard03,Nyman14,Nyman15b, Janhunen15}. The context-specific constraints increase the expressiveness of network models, however, this comes at the price of a much larger model space, which also has a more complicated topology. 

The above mentioned factors make structure learning an intricate task and Markov networks with context-specific independences have thus been nearly exclusively considered under the assumption of chordality of the underlying graph \citep{Nyman14,Janhunen15,Nyman15b}. By further constraining the permitted contexts, model scoring can be done analytically through the marginal likelihood \citep{Nyman14,Janhunen15}. An alternative method was recently presented in \citet{Nyman15b}, where model scoring was performed using penalized maximum likelihood estimation based on cyclical projection of joint probabilities. This method could also be applied to non-chordal networks, however, similar to the iterative proportional fitting algorithm \citep{Lauritzen96}, its computational complexity increases rapidly as a function of the number of parameter constraints induced by the model. 

In this article we generalize the Bayesian pseudo-likelihood scoring criterion introduced by \citet{PensarMPL} for Markov networks. This makes it possible to lift the restrictions made in previous works while still enabling efficient structure learning. The remainder of the article is structured as follows. In the next section we define context-specific independence for Markov networks and introduce the class of contextual Markov networks. In Section 3, the marginal pseudo-likelihood for contextual Markov networks is derived and its consistency is proven. Section 4 presents experiments with a synthetic dataset and a wide range of real datasets, demonstrating that contextual Markov networks lead to improved predictive and estimation accuracy compared with Markov networks. The final section provides some conclusions and identifies possibilities for further research.    

\section{Context-specific independence in Markov networks}

\subsection{Markov networks}
We consider a set of $d$ discrete random variables $X=X_V=\{X_{j}\}_{j\in V}$ where $V=\{1,\ldots,d\}$. Each variable $X_{j}$ takes values from a finite set of outcomes $\mathcal{X}_{j}=\{0,1,\ldots,r_{j}-1\}$ with cardinality $|\mathcal{X}_{j}|=r_j$. A subset of the variables, $S\subseteq V$, is denoted by $X_{S}=\{X_{j}\}_{j\in S}$ and the corresponding joint outcome space is specified by the Cartesian product $\mathcal{X}_{S}=\times_{j\in S}\mathcal{X}_{j}$. Occasionally, we omit the brackets from the subindex in order to improve readability. We use a lowercase letter $x_{S}$ to denote that the variables have been assigned a specific joint outcome in $\mathcal{X}_{S}$. Similarly, $p(x)$ is used as shorthand for the probability $p(X=x)$ while $p(X)$ refers to the distribution over $X$.  

A joint distribution over $X$ can be specified directly by the joint probabilities $\{p(x): x\in\mathcal{X}\}$. However, restricting the distributions to being strictly positive allows us to use the log-linear parameterization 
\begin{equation*}
\log p(x)=\sum_{A\subseteq V} \phi_A(x_A) 
\end{equation*}
where the $\phi$-terms are real-valued coordinate projection functions such that $\phi_A(x)=\phi_A(x_A)$ \citep{Whittaker90}. In order to avoid an over-parameterization, the functions are defined such that
\begin{equation}\label{eq:zero_rest}
\phi_{A}(x_A)=0 \text{ if }x_j=0 \text{ for any } j\in A. 
\end{equation}
The above restriction ensures a one-to-one correspondence between the joint probabilities and the functions. The latter can be determined from the former by solving a triangular system of linear equations. A convenient property of the log-linear expansion is that restrictions related to conditional independence correspond to setting $\phi$-terms to zero. We use $X_A\perp X_B \mid X_C$ to denote that $X_A$ is conditionally independent of $X_B$ given $X_C$, that is $p(X_A\mid X_B,X_C)=p(X_A\mid X_C)$. Now, if $(V_1,V_2,V_3)$ is a partition of $V$, then 
\begin{equation}\label{eq:ci_log_lin}
X_{V_1}\perp X_{V_2} \mid X_{V_3} \Leftrightarrow \phi_{A}(\cdot)=0 \text{ when } A\cap V_1\not= \varnothing \text{ and } A\cap V_2\not= \varnothing.
\end{equation}

A Markov network (MN) over $X$ is a probabilistic graphical model that compactly represents a joint distribution over the variables by exploiting statements of conditional independence. The dependence structure over the $d$ variables is specified by an undirected graph $G=(V,E)$ where the nodes (or vertices) $V=\{1,\ldots,d\}$ correspond to the indices of the variables and the edges $E\subseteq \{ V\times V\}$ represent dependences among the variables. As is common in graphical model literature, we will use the terms node and variable interchangeably throughout this article. A node $j$ is a neighbor of $i$ (and vice versa) if $\{i,j\}\in E$. The set of all neighbors of a node $j$ is called the Markov blanket of the node, $mb(j)=\{ i\in V:\{i,j\}\in E \}$. We denote the set of common neighbors to two nodes $i$ and $j$ by $cn(i,j)=mb(i)\cap mb(j)$. Finally, we will denote the corresponding set of common neighbors with the edge included by $\overbar{cn}(i,j)=cn(i,j)\cup\{ i,j \}$. 

Absence of edges in the graph of an MN encodes statements of conditional independence which can be characterized by the following Markov properties:
\begin{enumerate}
\item Pairwise Markov property: $X_{i}\perp X_{j}\mid X_{V\setminus \{ i,j \}}$ for all $\{i,j\}\not\in E$.
\item Local Markov property: $X_{i}\perp X_{V\setminus \{ mb(i)\cup i \}}\mid X_{mb(i)}$ for all $i\in V$.
\item Global Markov property: $X_{A}\perp X_{B}\mid X_{S}$ for all disjoint subsets $(A,B,S)$ of $V$ such that $S$ separates $A$ from $B$. 
\end{enumerate}
Although the strength of the above properties differ in general, they are proven to be equivalent under the current assumption of positivity of the joint distribution \citep[][Theorem 3.7]{Lauritzen96}.

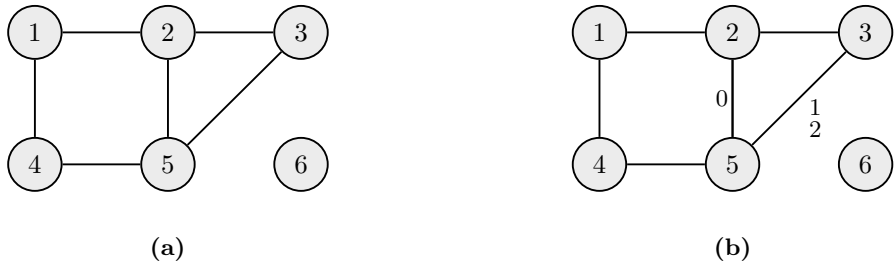
\begin{figure}
\begin{subfigure}{0.5\textwidth}
\begin{center}
\begin{tikzpicture}
\node[at={(0,0)},circle,draw,fill=node_color,line width=0.25mm,minimum size=20pt](4){4};
\node[at={(1.75,0)},circle,draw,fill=node_color,line width=0.25mm,minimum size=20pt](5){5};
\node[at={(3.5,0)},circle,draw,fill=node_color,line width=0.25mm,minimum size=20pt](6){6};
\node[at={(0,1.75)},circle,draw,fill=node_color,line width=0.25mm,minimum size=20pt](1){1};
\node[at={(1.75,1.75)},circle,draw,fill=node_color,line width=0.25mm,minimum size=20pt](2){2};
\node[at={(3.5,1.75)},circle,draw,fill=node_color,line width=0.25mm,minimum size=20pt](3){3};
\path (1) edge[line width=0.25mm] (2);
\path (2) edge[line width=0.25mm] (3);
\path (1) edge[line width=0.25mm] (4);
\path (4) edge[line width=0.25mm] (5);
\path (2) edge[line width=0.25mm] (5);
\path (3) edge[line width=0.25mm] (5);
\end{tikzpicture}
\end{center}
\caption{\label{fig:ex_ug_a}}
\end{subfigure}
\begin{subfigure}{0.5\textwidth}
\begin{center}
\begin{tikzpicture}
\node[at={(0,0)},circle,draw,fill=node_color,line width=0.25mm,minimum size=20pt](4){4};
\node[at={(1.75,0)},circle,draw,fill=node_color,line width=0.25mm,minimum size=20pt](5){5};
\node[at={(3.5,0)},circle,draw,fill=node_color,line width=0.25mm,minimum size=20pt](6){6};
\node[at={(0,1.75)},circle,draw,fill=node_color,line width=0.25mm,minimum size=20pt](1){1};
\node[at={(1.75,1.75)},circle,draw,fill=node_color,line width=0.25mm,minimum size=20pt](2){2};
\node[at={(3.5,1.75)},circle,draw,fill=node_color,line width=0.25mm,minimum size=20pt](3){3};
\path (1) edge[line width=0.25mm] (2);
\path (2) edge[line width=0.25mm] (3);
\path (1) edge[line width=0.25mm] (4);
\path (4) edge[line width=0.25mm] (5);
\path (2) edge[line width=0.25mm] node[left=-2] {\small{$0$}} (5);
\path (3) edge[line width=0.25mm] node[below right=-5] {\small{$\stack{1}{2}$}} (5);
\path (2) edge[line width=0.25mm] (5);
\end{tikzpicture}
\end{center}
\caption{\label{fig:ex_ug_b}}
\end{subfigure}
\caption{A (a) traditional and (b) labeled undirected graph over six nodes.\label{fig:ex_ug}}
\end{figure}

To fully specify an MN, one must define a probability distribution $P$ over the variables. The distribution must satisfy the conditional independence restrictions imposed by the graph. If the distribution does not satisfy any additional independences, not conveyed by the graph, the distribution is said to be faithful to the graph. Under the log-linear parameterization, the graph-induced restrictions are accounted for by setting certain $\phi$-terms to zero. More specifically, by combining the property stated in \eqref{eq:ci_log_lin} and the pairwise Markov property, we reach the conclusion
\begin{equation}\label{eq:g_loglin_rest}
\{i,j\}\not\in E \Rightarrow \phi_{A}(\cdot)=0 \text{ if } \{i,j\}\subseteq A.
\end{equation}
In other words, all $\phi$-terms covering pairs of nodes not connected by an edge are equal to zero. Consequently, the log-linear parameterization induced by an undirected graph is in general hierarchical in the sense that if $\phi_A(\cdot)=0$ then $\phi_B(\cdot)=0$ for all $B\supseteq A$. An example of an undirected six-node graph is shown in Figure \ref{fig:ex_ug}(a). The corresponding MN has the log-linear expansion
\begin{equation}\label{eq:log_lin_para}
\begin{aligned}
\log p(x)&=\phi_{\varnothing}+\phi_{1}(x)+\phi_{2}(x)+\phi_{3}(x)+\phi_{4}(x)+\phi_{5}(x)+\phi_{6}(x)\\
&+\phi_{1,2}(x)+\phi_{1,4}(x)+\phi_{2,3}(x)+\phi_{2,5}(x)+\phi_{3,5}(x)+\phi_{4,5}(x)+\phi_{2,3,5}(x)
\end{aligned}
\end{equation}
Note that no $\phi$-term is subscripted by pairs of nodes that are not in the edge set.

\subsection{Contextual Markov networks}
It has previously been noticed that conditional independence alone may be unnecessarily stringent for modeling real-world phenomena \citep{Boutilier96,Friedman96,Geiger96,Chickering97,Poole03,Corander03,Hojsgaard03,Nyman14,Pensar15,Nyman15b}. For this reason, the notion of context-specific independence (CSI) has been proposed as a tool to relax the restrictions of graphical models while still providing a sound independence-based interpretation. The class of CSI is essentially a generalization of conditional independence that only holds in a certain context, specified by the conditioning variables. The concept of CSI was formalized by \citet{Boutilier96} for the purpose of refining the conditional probability tables of Bayesian networks \citep{Boutilier96,Friedman96,Poole03,Pensar15}, however, it has also been studied as a means to refine Markov networks \citep{Corander03,Hojsgaard03,Nyman14,Nyman15b}.

\begin{definition} Context-Specific Independence \\
\normalfont Let $X=\{X_{1},\ldots,X_{d}\}$ be a set of stochastic variables where $V=\{1,\ldots d\}$ and let $A$, $B$, $C$, $S$ be four disjoint subsets of $V$. $X_{A}$ is contextually independent of $X_{B}$ given $X_{S}$ and the context $X_{C}=x_{C}$ if  
\[
p(x_{A}\mid x_{B},x_{C},x_{S})=p(x_{A}\mid x_{C},x_{S})
\]
holds for all $(x_{A},x_{B},x_{S})\in\mathcal{X}_{A}\times\mathcal{X}_{B}\times\mathcal{X}_{S}$ whenever $p(x_{B},x_{C},x_{S})>0$. This will be denoted by 
\[
X_{A}\perp X_{B}\mid x_{C},X_{S}.
\]
\end{definition}

To incorporate CSI in Markov networks in a general setting, we introduce the concept of contextual Markov networks (CMN).
\begin{definition} Contextual Markov network structure \label{def:cmn_struct}\\
\normalfont The dependence structure of a contextual Markov network is a pair $(G,\mathcal{C})$ consisting of an undirected graph $G=(V,E)$, whose nodes represent random variables $X_1,\ldots,X_d$, and a set of contexts $\mathcal{C}=\{ \scal(i,j) \}_{\{ i,j \}\in E}$ where $\scal(i,j)$ is the edge context of edge $\{ i,j \}$. An edge context $\scal(i,j)$ is a subset of the outcome space of the common neighbors $cn(i,j)$. The pair $(G,\mathcal{C})$ encodes a dependence structure over the variables according to:
\begin{enumerate}
\item $X_1,\ldots,X_d$ satisfy the Markov properties of the undirected graph.
\item For each $\{ i,j \}\in E$ and $x_{cn(i,j)} \in \scal(i,j):$ $X_i \perp X_j \mid x_{cn(i,j)},X_{V\setminus \overbar{cn}(i,j)}$.
\end{enumerate}
\end{definition} 
An edge context represents a set of situations, specified by the common neighbors, where the direct influence implied by the edge is cut off. Note that an edge may have a non-empty edge context only if the edge nodes have at least one neighbor in common, or equivalently, the edge is part of a clique of size three or larger. If the nodes of an edge have more than one common neighbor we assume an ascending ordering of the neighbors according to the node indices.

\begin{figure}
\begin{center}
\begin{tikzpicture}
\node[at={(0,0)},circle,draw,fill=node_color,line width=0.25mm,minimum size=20pt](3){3};
\node[at={(1.75,0)},circle,draw,fill=node_color,line width=0.25mm,minimum size=20pt](4){4};
\node[at={(0,1.75)},circle,draw,fill=node_color,line width=0.25mm,minimum size=20pt](1){1};
\node[at={(1.75,1.75)},circle,draw,fill=node_color,line width=0.25mm,minimum size=20pt](2){2};
\path (1) edge[line width=0.25mm] (2);
\path (1) edge[line width=0.25mm] node[left=-6]{\small{$\stack{01}{10}$}} (3);
\path (1) edge[line width=0.25mm] (4);
\path (2) edge[line width=0.25mm] (3);
\path (2) edge[line width=0.25mm] node[right=-1]{\small{$1*$}} (4);
\path (3) edge[line width=0.25mm] (4);
\end{tikzpicture}
\end{center}
\caption{A labeled undirected graph over four binary variables.\label{fig:ex_label}}
\end{figure}
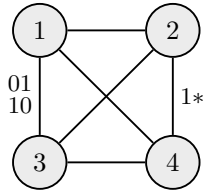

To visualize the graph-context structure as a single entity, we assign labels to edges with non-empty contexts as originally proposed by \citet{Corander03}, who introduced the class of labeled graphical models. As an example of how to interpret labels, consider Figure \ref{fig:ex_label} which depicts a labeled graph over four binary variables. For illustrative purposes, a label stacks the context configurations instead of drawing them as an actual set, that is, the label on edge $\{1,3\}$ corresponds to the edge context $\scal(1,3)=\{(0,1),(1,0)\}$. As established earlier, an ascending ordering of the common neighbors is used. For this particular edge context, it means that the first position in a configuration contains a value of variable $X_2$ and the second position contains a value of $X_4$. Moreover, the star symbol is used to represent a collection of configurations obtained by allowing the variable corresponding to the star position to take on any value, for example, the label on edge $(2,4)$ represents the edge context $\scal(2,4)=1\times \mathcal{X}_3 = 1 \times \{0,1\}=\{(1,0),(1,1)\}$.

The definition of edge context differs from previous works in that it explicitly considers the direct influence of an edge by conditioning on the remaining variables rather then only on the common neighbors. This modification is due to an observation concerning non-chordal graphs made in \citet{Nyman15b}. To illustrate the phenomenon, consider the non-chordal labeled graph in Figure \ref{fig:ex_ug}(b). The label on edge $\{ 2,5 \}$ does not imply that 
\[
X_2 \perp X_5\mid X_3=0
\]
due to the chordless loop $2-1-4-5$. This type of situations cannot occur in chordal graphs. However, to maintain the soundness of an edge context when considering non-chordal graphs, the above definition has been formulated to make a statement explicitly about the direct dependence between the edge variables, that is
\[
X_2 \perp X_5\mid X_3=0,X_{1,4,6}.
\]

\begin{definition} Contextual Markov network \label{def:cmn_hmm}\\
\normalfont A contextual Markov network is a triple $(G,\mathcal{C},P)$ where $P$ denotes a distribution over $X_V$ that satisfies the dependence structure encoded by the contextual Markov network structure $(G,\mathcal{C})$.
\end{definition} 
The class of CMNs is essentially the same as the class of labeled graphical models, except for the revised definition of edge context, and subsumes the classes of stratified graphical models \citep{Nyman14,Nyman15b} which are restricted to chordal graphs. 

The reason for specifying an edge context by the common neighbors has been shown to be a natural condition \citep[see][Theorem 2]{Nyman15b}. In the following result, we further demonstrate the reason for this from an independence-based perspective.
\begin{prop}\label{thm:cn}
Let $X_{V}$ be a set of variables satisfying the Markov properties of the undirected graph $G=(V,E)$. Furthermore, let $i,j,k\in V$ such that $\{i,j\}\in E$ and $k\not\in cn(i,j)$. Under the Markov properties of $G$, the context-specific independence statement
\[
X_{i} \perp X_{j} \mid x_{cn(i,j)},x_k,X_{V\setminus \{ \overbar{cn}(i,j)\cup k\} }
\]
is equivalent to
\[
X_{i} \perp X_{j} \mid x_{cn(i,j)},X_{V\setminus \overbar{cn}(i,j) }.
\]
\end{prop}
\begin{proof}\let\qed\relax
See appendix.
\end{proof}
In other words, including variables that are not common neighbors into the set that specifies the edge context would not increase the generality of the models. Obviously, it would be possible to let an edge context be specified by a subset of the common neighbors, however, this would not increase the generality of the models since such a CSI statement is equivalent to a collection of CSIs of the type in Definition \ref{def:cmn_struct}. For example, say that the context of edge $\{2,4\}$ in the graph in Figure \ref{fig:ex_label} would be $\scal(2,4)=\{1\}$ such that it would be specified by node $1\subseteq cn(2,4)$. This would imply that
\[
 X_2 \perp X_4 \mid X_1=1,X_3,
\]
which clearly is equivalent to the two statements
\[
X_2 \perp X_4 \mid X_1=1,X_3=0\text{ and } X_2 \perp X_4 \mid X_1=1,X_3=1,
\]
which correspond to the original proper edge context $\scal(2,4)=\{(1,0),(1,1)\}$.

It has previously been shown that local CSIs similar to the type considered here result in linear restrictions on the parameters in the log-linear parameterization \citep{Corander03,Nyman15b}. In line with previous works, we show that this also holds for CMNs.
\begin{prop}\label{thm:ec_loglin_rest}
An element in an edge context, $x_{cn(i,j)}\in \scal(i,j)$, imposes the following $(r_{i}-1)(r_{j}-1)$ linear restrictions on the log-linear parameters:
\[
\sum_{A\subseteq cn(i,j)}\phi_{A\cup \{ i,j \}}(x_{A},x'_{i},x'_{j})=0 \text{ for all } x'_{\{i,j\}}\in \{1,\ldots,r_{i}-1\}\times\{1,\ldots,r_{j}-1\}
\]
where $\phi_{A\cup \{ i,j \}}(x_{A},x'_{i},x'_{j})=0$ if $x_{k}=0$ for any $k\in A$. 
\end{prop}
\begin{proof}\let\qed\relax
See appendix.
\end{proof}
Note that $\phi_{A\cup \{ i,j \}}(x_{A},x'_{i},x'_{j})=0$ in the above definition if $k,l\in A$ and $\{k,l\}\not\in E$ due to restriction \eqref{eq:g_loglin_rest}. To provide an example of how the above proposition can be used in practice, consider the labeled graph in Figure \ref{fig:ex_ug}(b). Assuming that all the variables have an outcome space equal to $\{ 0,1,2 \}$, the edge contexts induce the following restrictions on the $\phi$-functions in \eqref{eq:log_lin_para}:
\begin{center}
\begin{tabular}{l l l}
$0\in \scal(2,5):$\vspace{0.2cm} & $1\in \scal(3,5):$ & $2\in \scal(3,5):$ \\
$\phi_{2,5}(1,1)=0$\hspace{1cm} & $\phi_{3,5}(1,1)+\phi_{2,3,5}(1,1,1)=0$\hspace{1cm} & $\phi_{3,5}(1,1)+\phi_{2,3,5}(2,1,1)=0$ \\
$\phi_{2,5}(1,2)=0$ & $\phi_{3,5}(1,2)+\phi_{2,3,5}(1,1,2)=0$ & $\phi_{3,5}(1,2)+\phi_{2,3,5}(2,1,2)=0$ \\
$\phi_{2,5}(2,1)=0$ & $\phi_{3,5}(2,1)+\phi_{2,3,5}(1,2,1)=0$ & $\phi_{3,5}(2,1)+\phi_{2,3,5}(2,2,1)=0$ \\
$\phi_{2,5}(2,2)=0$ & $\phi_{3,5}(2,2)+\phi_{2,3,5}(1,2,2)=0$ & $\phi_{3,5}(2,2)+\phi_{2,3,5}(2,2,2)=0$ \\
\end{tabular}
\end{center}

In contrast to traditional MNs, the log-linear parameterization of a CMN is in general non-hierarchical \citep{Nyman15b}. Still, from Proposition \ref{thm:ec_loglin_rest} it is clear that the distribution of a CMN belongs to the collection of distributions of an MN with the same underlying graph. Consequently, a CMN indeed satisfies the Markov properties associated with its undirected graph making it a natural CSI-based extension of traditional MNs. 

From previous works \citep{Corander03,Nyman14}, it is obvious that distinct CMN structures may represent the same dependence structure. To avoid such an overlap, \citet{Corander03} introduced two conditions, \emph{maximality} and \emph{regularity}, which together guarantee that two distinct structures do not imply the same set of restrictions. The maximality condition ensures that no element can be added to an edge context without implying additional restrictions on the log-linear parameters. The regularity condition ensures that no edge can be completely removed by its edge context. Consequently, an edge context in a regular maximal CMN must be a strict subset of the outcome space of the common neighbors. Without loss of generality, it is sufficient to only consider regular maximal CMN structures \citep[see][for more details]{Corander03,Nyman14}.

\section{Marginal pseudo-likelihood learning of contextual Markov network structures}
One of the main reasons for restricting the context-specific models in \citep{Nyman14,Nyman15b} to chordal graphs is due to the complexity of learning the model structures from data. For the model class introduced in \citet{Nyman14}, there is a closed-form expression of the marginal likelihood making the model selection tractable for large systems, however, this necessitates fairly strict constraints on the CSIs allowed in the models. In \citet{Nyman15b} the restrictions were lifted making the marginal likelihood intractable to evaluate. The marginal likelihood was therefore replaced by the asymptotically equivalent Bayesian information criterion (BIC) by \citet{Schwarz78}. Still, the task of calculating the maximum likelihood estimates (MLEs) limits the applicability of the structure learning method to small-scale systems.

In \citet{PensarMPL}, the marginal pseudo-likelihood (MPL) score was introduced as a criterion for large-scale structure learning of general MNs. It was shown to perform very well against competing methods on both synthetic and real-world networks. In terms of traditional MNs, the aim is to recover the dependence structure which is specified by an undirected graph. In this section we show how the scope of the MPL can be extended to also handle the general class of CMNs whose dependence structure is specified by an undirected graph and an associated context.

\subsection{Marginal pseudo-likelihood for Markov networks}
By structure learning, we refer to the process of inferring the dependence structure of a model from a set of data assumed to be generated by a model in the considered model class. A dataset $\mathbf{x}$ refers to a collection of $n$ i.i.d. complete joint observations over the $d$ variables. For derivation purposes, we introduce the graph-specific parameters 
\begin{equation}\label{eq:para}
\theta_{ijl}=p(x_{j}^{(i)}\mid x_{mb(j)}^{(l)})
\end{equation}
where the upper indices $i$ and $l$ represent specific outcomes of variable $j$ and its Markov blanket $mb(j)$. Similarly, we denote the counts of the corresponding configurations in $\mathbf{x}$ by $n_{ijl}$. 

In the Bayesian framework, a graph is scored by its posterior probability given the data,
\[
p(G\mid \mathbf{x})\propto p(\mathbf{x}\mid G)\cdot p(G), 
\]
where $p(\mathbf{x}\mid G)$ is the marginal likelihood and $p(G)$ is the prior probability of the graph. The marginal likelihood is the key component of the Bayesian score and it is evaluated as
\begin{equation}\label{eq:marg_lh}
p(\mathbf{x}\mid G)=\int p(\mathbf{x}\mid \theta_{G})\cdot f(\theta_{G})d\theta_{G}, 
\end{equation}
where $\theta_G$ represents the set of parameters specifying a distribution imposed by $G$, $p(\mathbf{x}\mid \theta_{G})$ is the probability of the data under the given parameters (known as the likelihood of the parameters), and $f(\theta_{G})$ is a prior distribution over the possible parameter values. The marginal likelihood implicitly penalizes overly dense graphs through its parameter prior. In contrast to MNs for which the marginal likelihood is in general intractable, the Bayesian score has become the most popular choice when learning the related class of Bayesian networks for which the likelihood function factorizes into separate node-wise conditional likelihoods according to a directed acyclic graph (DAG).

Due to the intractability of the marginal likelihood for non-chordal MNs, \citet{PensarMPL} introduced the MPL as a computationally efficient alternative score. The score is based on the pseudo-likelihood function \citep{Besag75} which approximates the likelihood function by factorizing it into node-wise conditional likelihood functions. In particular, for a given graph, the pseudo-likelihood function is simplified to
\[
\hat p (\mathbf{x} \mid \theta_G)=\prod_{j=1}^{d} p (\mathbf{x}_j \mid \mathbf{x}_{mb(j)}, \theta_G)=\prod_{j=1}^{d} \prod_{l=1}^{q_j} \prod_{i=1}^{r_j} \theta_{ijl}^{n_{ijl}}
\]
since each variable is independent of the rest of the network given its Markov blanket (local Markov property). The MPL is now obtained by replacing the likelihood function in \eqref{eq:marg_lh} with the pseudo-likelihood function. Under certain assumptions regarding parameter independence \citep[see][]{PensarMPL}, the parameter prior can be factorized in a similar fashion. By further assuming 
\begin{equation*}
(\theta_{1jl},\ldots,\theta_{r_{j}jl})\sim \text{Dirichlet}(\alpha_{1jl},\ldots,\alpha_{r_{j}jl})\text{ for all } j=1,\ldots,d\text{ and }l=1,\ldots,q_j,
\end{equation*}
the MPL can be evaluated by the closed-form expression
\begin{equation}\label{eq:mpl}
\hat{p}(\mathbf{x}\mid G)=\prod_{j=1}^{d}\prod_{l=1}^{q_j}\frac{\Gamma(\alpha_{jl})}{\Gamma(n_{jl}+\alpha_{jl})}\prod_{i=1}^{r_j}\frac{\Gamma(n_{ijl}+\alpha_{ijl})}{\Gamma(\alpha_{ijl})}
\end{equation}
where $n_{jl}=\sum_{i=1}^{r_j}n_{ijl}$ and $\alpha_{jl}=\sum_{i=1}^{r_j}\alpha_{ijl}$. In practice, the logarithm of the formula is used since it is computationally more manageable. We specify the hyperparameters by setting $\alpha_{ijl}=1/2$, which results in Jeffreys' prior for the multinomial distribution.

\begin{figure}
\begin{center}
\begin{tikzpicture}
\node[at={(0,-1.75)},circle,draw,line width=0.25mm,minimum size=20pt](4){4};
\node[at={(0,0)},circle,draw,fill=node_color,line width=0.25mm,minimum size=20pt](1){1};
\node[at={(1.75,0)},circle,draw,line width=0.25mm,minimum size=20pt](2){2};
\draw[line width=0.25mm,decoration={markings,mark=at position 1 with {\arrow[scale=1.5]{stealth}}},postaction={decorate}] (2) -- (1);
\draw[line width=0.25mm,decoration={markings,mark=at position 1 with {\arrow[scale=1.5]{stealth}}},postaction={decorate}] (4) -- (1);
\node[at={(4.75,-1.75)},circle,draw,line width=0.25mm,minimum size=20pt](5){5};
\node[at={(3,0)},circle,draw,line width=0.25mm,minimum size=20pt](1){1};
\node[at={(4.75,0)},circle,draw,fill=node_color,line width=0.25mm,minimum size=20pt](2){2};
\node[at={(6.5,0)},circle,draw,line width=0.25mm,minimum size=20pt](3){3};
\draw[line width=0.25mm,decoration={markings,mark=at position 1 with {\arrow[scale=1.5]{stealth}}},postaction={decorate}] (1) -- (2);
\draw[line width=0.25mm,decoration={markings,mark=at position 1 with {\arrow[scale=1.5]{stealth}}},postaction={decorate}] (3) -- (2);
\draw[line width=0.25mm,decoration={markings,mark=at position 1 with {\arrow[scale=1.5]{stealth}}},postaction={decorate}] (5) -- (2);
\node[at={(7.75,-1.75)},circle,draw,line width=0.25mm,minimum size=20pt](5){5};
\node[at={(7.75,0)},circle,draw,line width=0.25mm,minimum size=20pt](2){2};
\node[at={(9.5,0)},circle,draw,fill=node_color,line width=0.25mm,minimum size=20pt](3){3};
\draw[line width=0.25mm,decoration={markings,mark=at position 1 with {\arrow[scale=1.5]{stealth}}},postaction={decorate}] (2) -- (3);
\draw[line width=0.25mm,decoration={markings,mark=at position 1 with {\arrow[scale=1.5]{stealth}}},postaction={decorate}] (5) -- (3);
\node[at={(0,-4.75)},circle,draw,fill=node_color,line width=0.25mm,minimum size=20pt](4){4};
\node[at={(0,-3)},circle,draw,line width=0.25mm,minimum size=20pt](1){1};
\node[at={(1.75,-4.75)},circle,draw,line width=0.25mm,minimum size=20pt](5){5};
\draw[line width=0.25mm,decoration={markings,mark=at position 1 with {\arrow[scale=1.5]{stealth}}},postaction={decorate}] (1) -- (4);
\draw[line width=0.25mm,decoration={markings,mark=at position 1 with {\arrow[scale=1.5]{stealth}}},postaction={decorate}] (5) -- (4);
\node[at={(3,-4.75)},circle,draw,line width=0.25mm,minimum size=20pt](4){4};
\node[at={(4.75,-3)},circle,draw,line width=0.25mm,minimum size=20pt](2){2};
\node[at={(6.5,-3)},circle,draw,line width=0.25mm,minimum size=20pt](3){3};
\node[at={(4.75,-4.75)},circle,draw,fill=node_color,line width=0.25mm,minimum size=20pt](5){5};
\draw[line width=0.25mm,decoration={markings,mark=at position 1 with {\arrow[scale=1.5]{stealth}}},postaction={decorate}] (4) -- (5);
\draw[line width=0.25mm,decoration={markings,mark=at position 1 with {\arrow[scale=1.5]{stealth}}},postaction={decorate}] (2) -- (5);
\draw[line width=0.25mm,decoration={markings,mark=at position 1 with {\arrow[scale=1.5]{stealth}}},postaction={decorate}] (3) -- (5);
\node[at={(9.5,-4.75)},circle,draw,fill=node_color,line width=0.25mm,minimum size=20pt](6){6};
\end{tikzpicture}
\end{center}
\caption{Node-specific DAGs illustrating the factorization of MPL for the graph in Figure \ref{fig:ex_ug}(a). \label{fig:ex_ug_mpl}}
\end{figure}
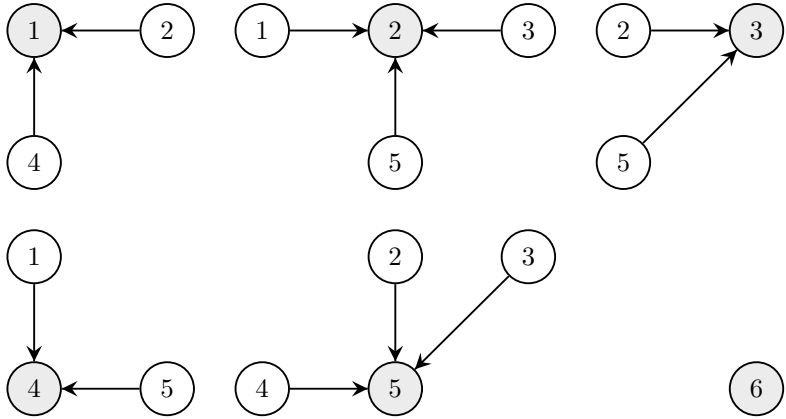

Similar to the marginal likelihood of a Bayesian network, the MPL factorizes into node-wise marginal conditional likelihoods: 
\begin{equation}\label{eq:mpl_fac}
\hat{p}(\mathbf{x}\mid G)=\prod_{j=1}^{d} p(\mathbf{x}_j \mid \mathbf{x}_{mb(j)}).
\end{equation}
For example, the MPL of the undirected graph in Figure \ref{fig:ex_ug}(a) follows a node-wise factorization illustrated by the DAGs in Figure \ref{fig:ex_ug_mpl}:
\[
\hat{p}(\mathbf{x}\mid G)=p(\mathbf{x}_1 \mid \mathbf{x}_{2,4})p(\mathbf{x}_2 \mid \mathbf{x}_{1,3,5})p(\mathbf{x}_3 \mid \mathbf{x}_{2,5})p(\mathbf{x}_4 \mid \mathbf{x}_{1,5})p(\mathbf{x}_5 \mid \mathbf{x}_{2,3,4})p(\mathbf{x}_6).
\]
The set of parents of a node $j$, which is defined as all nodes from which there are directed edges towards $j$, is the Markov blanket $mb(j)$ of the node in the undirected graph. The Markov blankets of a graph are mutually consistent in the sense that $i\in mb(j)$ iff $j\in mb(i)$. In the local DAGs, this is reflected by $i$ being a parent to $j$ iff $j$ is a parent to $i$. Consequently, when combining the collection of local DAGs associated with the MPL, the result is not a DAG in itself but rather a bi-directional dependency network \citep{Heckerman01}.

In addition to the MPL, we also need to specify a graph prior $p(G)$, through which it is possible to incorporate any prior belief concerning the graph structure. In order to preserve the decomposability of the overall score, the prior must follow a similar factorization as the MPL \citep[see][]{PensarMPL}.

\subsection{Marginal pseudo-likelihood for contextual Markov networks}
In this section we show how the MPL can be modified to also take the context structure into account. More specifically, we show that the CSIs of a CMN can be taken into account in a similar fashion as CSI has been exploited in learning of Bayesian networks \citep{Friedman96,Pensar15}. The key innovation lies in the observation that the CSIs in Definition \ref{def:cmn_struct} can be recast to be consistent with the structure of MPL. More specifically, under the current assumptions and a given graph, the local Markov property ensures that 
\begin{equation}\label{eq:mb_csi}
X_i \perp X_j \mid x_{cn(i,j)},X_{V\setminus \overbar{cn}(i,j)} \Leftrightarrow
\begin{aligned}
&X_i \perp X_j \mid x_{cn(i,j)},X_{mb(i) \setminus \{ cn(i,j) \cup j  \}}\\
&X_i \perp X_j \mid x_{cn(i,j)},X_{mb(j) \setminus \{ cn(i,j) \cup i \}}
\end{aligned},
\end{equation}
where the CSIs on the right hand side imply that some of the conditional distributions of a node given its Markov blanket are assumed identical. This is essentially equivalent to including local CSIs \citep[][Definition 3]{Pensar15} in learning of Bayesian networks. 

Constraints of the above type are thus straightforward to incorporate into the evaluation of the MPL. In practice, the outcome space of the conditioning variables is partitioned into classes with invariant conditional distributions for outcomes assigned to the same class. Consequently, the corresponding conditional distribution parameters \eqref{eq:para} are defined over classes rather than distinct configurations \citep{Friedman96,Chickering97,Pensar15}. Following the same approach as in the previous section, we reach the same closed-form expression \eqref{eq:mpl}, however, the $l$-index now runs over classes which are specified by the edge contexts. Similarly, the counts $n_{ijl}$'s are now defined in terms of classes meaning that distinct Markov blanket configurations within the same class contribute to the same class-specific count.

\begin{figure}
\begin{center}
\begin{tikzpicture}
\node[at={(0,0)}](111){$\{(0,0,0)\}$};
\node[at={(0,-0.45)}](000){$\{(0,0,1)\}$};
\node[at={(0,-0.9)}](000){$\{(0,0,2)\}$};
\node[at={(0,-1.35)}](000){$\{(0,1,0)\}$};
\node[at={(0,-1.8)}](000){$\{(0,1,1)\}$};
\node[at={(0,-2.25)}](000){$\{(0,1,2)\}$};
\node[at={(0,-2.7)}](000){$\{(0,2,0)\}$};
\node[at={(0,-3.15)}](000){$\{(0,2,1)\}$};
\node[at={(0,-3.6)}](000){$\{(0,2,2)\}$};
\node[at={(0,-4.05)}](1){$\{ (1,0,0)\}$};
\node[at={(0,-4.5)}](5){$\{ (1,0,1)\}$};
\node[at={(0,-4.95)}](9){$\{ (1,0,2)\}$};
\node[at={(0,-5.4)}](2){$\{ (1,1,0)\}$};
\node[at={(0,-5.85)}](6){$\{ (1,1,1)\}$};
\node[at={(0,-6.3)}](10){$\{ (1,1,2)\}$};
\node[at={(0,-6.75)}](3){$\{ (1,2,0)\}$};
\node[at={(0,-7.2)}](7){$\{ (1,2,1)\}$};
\node[at={(0,-7.65)}](11){$\{ (1,2,2)\}$};
\node[at={(0,-8.1)}](000){$\{(2,0,0)\}$};
\node[at={(0,-8.55)}](000){$\{(2,0,1)\}$};
\node[at={(0,-9)}](000){$\{(2,0,2)\}$};
\node[at={(0,-9.45)}](000){$\{(2,1,0)\}$};
\node[at={(0,-9.9)}](000){$\{(2,1,1)\}$};
\node[at={(0,-10.35)}](000){$\{(2,1,2)\}$};
\node[at={(0,-10.8)}](000){$\{(2,2,0)\}$};
\node[at={(0,-11.25)}](000){$\{(2,2,1)\}$};
\node[at={(0,-11.7)}](000){$\{(2,2,2)\}$};
\node[at={(4,0)}](222){$\{(0,0,0)\}$};
\node[at={(4,-0.45)}](000){$\{(0,0,1)\}$};
\node[at={(4,-0.9)}](000){$\{(0,0,2)\}$};
\node[at={(4,-1.35)}](000){$\{(0,1,0)\}$};
\node[at={(4,-1.8)}](000){$\{(0,1,1)\}$};
\node[at={(4,-2.25)}](000){$\{(0,1,2)\}$};
\node[at={(4,-2.7)}](000){$\{(0,2,0)\}$};
\node[at={(4,-3.15)}](000){$\{(0,2,1)\}$};
\node[at={(4,-3.6)}](000){$\{(0,2,2)\}$};
\node[at={(4,-4.5)}](4){$\begin{Bmatrix}(1,0,0) \\  (1,1,0) \\  (1,2,0)\end{Bmatrix}$};
\node[at={(4,-5.85)}](8){$\begin{Bmatrix} (1,0,1) \\  (1,1,1) \\  (1,2,1)\end{Bmatrix}$};
\node[at={(4,-7.2)}](12){$\begin{Bmatrix} (1,0,2) \\  (1,1,2) \\  (1,2,2)\end{Bmatrix}$};
\node[at={(4,-8.1)}](13){$\{(2,0,0)\}$};
\node[at={(4,-8.55)}](14){$\{(2,0,1)\}$};
\node[at={(4,-9)}](15){$\{(2,0,2)\}$};
\node[at={(4,-9.45)}](17){$\{(2,1,0)\}$};
\node[at={(4,-9.9)}](18){$\{(2,1,1)\}$};
\node[at={(4,-10.35)}](19){$\{(2,1,2)\}$};
\node[at={(4,-10.8)}](21){$\{(2,2,0)\}$};
\node[at={(4,-11.25)}](22){$\{(2,2,1)\}$};
\node[at={(4,-11.7)}](23){$\{(2,2,2)\}$};
\path (1.east) edge[line width=0.25mm] (4.west);
\path (2.east) edge[line width=0.25mm] (4.west);
\path (3.east) edge[line width=0.25mm] (4.west);
\path (5.east) edge[line width=0.25mm] (8.west);
\path (6.east) edge[line width=0.25mm] (8.west);
\path (7.east) edge[line width=0.25mm] (8.west);
\path (9.east) edge[line width=0.25mm] (12.west);
\path (10.east) edge[line width=0.25mm] (12.west);
\path (11.east) edge[line width=0.25mm] (12.west);
\node[at={(8,0)}](25){$\{(0,0,0)\}$};
\node[at={(8,-0.45)}](26){$\{(0,0,1)\}$};
\node[at={(8,-0.9)}](27){$\{(0,0,2)\}$};
\node[at={(8,-1.35)}](000){$\{(0,1,0)\}$};
\node[at={(8,-1.8)}](000){$\{(0,1,1)\}$};
\node[at={(8,-2.25)}](000){$\{(0,1,2)\}$};
\node[at={(8,-2.7)}](000){$\{(0,2,0)\}$};
\node[at={(8,-3.15)}](000){$\{(0,2,1)\}$};
\node[at={(8,-3.6)}](000){$\{(0,2,2)\}$};
\node[at={(8,-4.5)}](4){$\begin{Bmatrix}(1,0,0) \\ (1,1,0) \\ (1,2,0)\end{Bmatrix}$};
\node[at={(8,-5.85)}](8){$\begin{Bmatrix}(1,0,1) \\ (1,1,1) \\ (1,2,1)\end{Bmatrix}$};
\node[at={(8,-7.2)}](12){$\begin{Bmatrix}(1,0,2) \\ (1,1,2) \\ (1,2,2)\end{Bmatrix}$};
\node[at={(8,-8.55)}](16){$\begin{Bmatrix} (2,0,0) \\  (2,1,0) \\  (2,2,0)\end{Bmatrix}$};
\node[at={(8,-9.9)}](20){$\begin{Bmatrix} (2,0,1) \\  (2,1,1) \\  (2,2,1)\end{Bmatrix}$};
\node[at={(8,-11.25)}](24){$\begin{Bmatrix} (2,0,2) \\  (2,1,2) \\  (2,2,2)\end{Bmatrix}$};
\path (13.east) edge[line width=0.25mm] (16.west);
\path (14.east) edge[line width=0.25mm] (20.west);
\path (15.east) edge[line width=0.25mm] (24.west);
\path (17.east) edge[line width=0.25mm] (16.west);
\path (18.east) edge[line width=0.25mm] (20.west);
\path (19.east) edge[line width=0.25mm] (24.west);
\path (21.east) edge[line width=0.25mm] (16.west);
\path (22.east) edge[line width=0.25mm] (20.west);
\path (23.east) edge[line width=0.25mm] (24.west);
\node[at={(12,-4.5)}](32){$\begin{Bmatrix}(0,0,1) \\ (1,0,1) \\ (1,1,1) \\ (1,2,1) \\ (2,0,1) \\ (2,1,1) \\ (2,2,1)\end{Bmatrix}$};
\node[at={(12,-1.35)}](28){$\begin{Bmatrix}(0,0,0) \\ (1,0,0) \\ (1,1,0) \\ (1,2,0) \\ (2,0,0) \\ (2,1,0) \\ (2,2,0)\end{Bmatrix}$};
\node[at={(12,-7.65)}](36){$\begin{Bmatrix}(0,0,2) \\ (1,0,2) \\ (1,1,2) \\ (1,2,2) \\ (2,0,2) \\ (2,1,2) \\ (2,2,2)\end{Bmatrix}$};
\node[at={(12,-9.45)}](000){$\{(0,1,0)\}$};
\node[at={(12,-9.9)}](000){$\{(0,1,1)\}$};
\node[at={(12,-10.35)}](000){$\{(0,1,2)\}$};
\node[at={(12,-10.8)}](000){$\{(0,2,0)\}$};
\node[at={(12,-11.25)}](000){$\{(0,2,1)\}$};
\node[at={(12,-11.7)}](000){$\{(0,2,2)\}$};
\path (25.east) edge[line width=0.25mm] (28.west);
\path (16.east) edge[line width=0.25mm] (28.west);
\path (4.east) edge[line width=0.25mm] (28.west);
\path (26.east) edge[line width=0.25mm] (32.west);
\path (20.east) edge[line width=0.25mm] (32.west);
\path (8.east) edge[line width=0.25mm] (32.west);
\path (27.east) edge[line width=0.25mm] (36.west);
\path (24.east) edge[line width=0.25mm] (36.west);
\path (12.east) edge[line width=0.25mm] (36.west);

\node[at={(0,0.5)}]{\small $X_2\hspace{0.05cm}X_3\hspace{0.05cm}X_4$};
\node[at={(4,0.5)}]{\small $X_2\hspace{0.05cm}X_3\hspace{0.05cm}X_4$};
\node[at={(8,0.5)}]{\small $X_2\hspace{0.05cm}X_3\hspace{0.05cm}X_4$};
\node[at={(12,0.5)}]{\small $X_2\hspace{0.05cm}X_3\hspace{0.05cm}X_4$};

\draw[->,line width=0.25mm] (111.north east) to[out=60,in=120] (222.north west);
\draw[->,line width=0.25mm] (222.north east) to[out=60,in=120] (25.north west);
\draw[->,line width=0.25mm] (25.north east) to[out=60,in=120] (28.north west);
\node[at={(2,1.4)}]{\small $X_5\perp X_3 \mid X_2=1,X_4$};
\node[at={(6,1.4)}]{\small $X_5\perp X_3 \mid X_2=2,X_4$};
\node[at={(10,1.4)}]{\small $X_5\perp X_2 \mid X_3=0,X_4$};
\end{tikzpicture}
\end{center}
\caption{A stepwise procedure for partitioning the outcome space of the Markov blanket of variable $X_5$ (Figure \ref{fig:ex_ug}(b)) according to the CSI restrictions implied by the labels. The curly brackets indicate the different classes. For example, in the first column (from the left) no CSI restriction has been imposed and each configuration belongs to a distinct class. When moving to the second column, the CSI above the curved arrow (between the first and second column) is imposed implying that certain configurations (indicated by the connecting lines) give rise to identical distributions and are thereby placed in the same class. This procedure is then repeated for the remaining CSIs, eventually resulting in the last column, which represents the final partition in which all edge contexts, $\scal (5)=\{ \scal(2,5), \scal(3,5) \}$, are accounted for. \label{fig:oc_part}}
\end{figure}
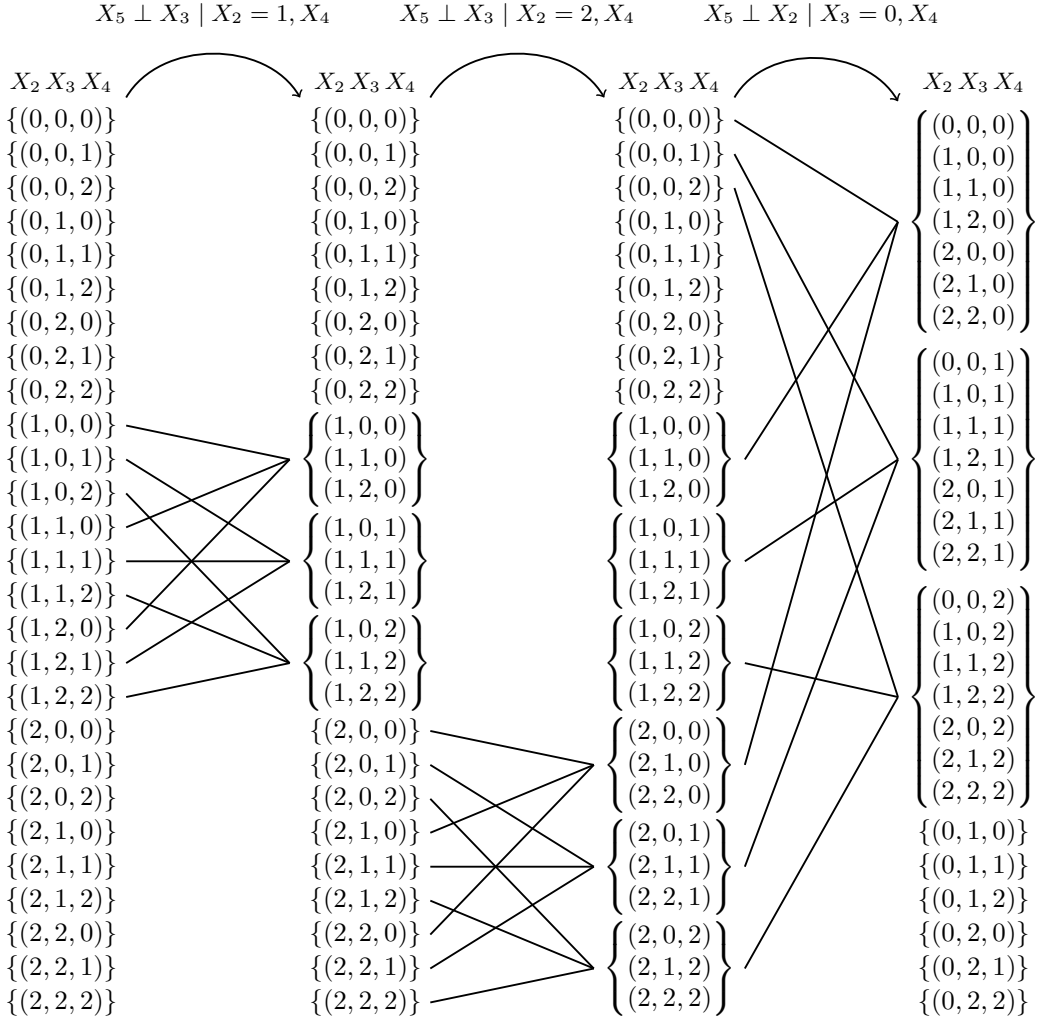

As an example, consider the CMN structure represented by the labeled graph in Figure \ref{fig:ex_ug}(b). The induced CSIs can be reformulated according to the equivalence relation in equation \eqref{eq:mb_csi}:
\begin{equation*}
\begin{aligned}
X_2 \perp X_5 \mid X_3=0,X_{1,4,6} & \Leftrightarrow
\begin{aligned}
&X_2 \perp X_5 \mid X_3=0,X_1\\
&X_5 \perp X_2 \mid X_3=0,X_4
\end{aligned},\\
X_3 \perp X_5 \mid X_2=1,X_{1,4,6} & \Leftrightarrow
\begin{aligned}
&X_3 \perp X_5 \mid X_2=1 \\
&X_5 \perp X_3 \mid X_2=1,X_4
\end{aligned},\\
X_3 \perp X_5 \mid X_2=2,X_{1,4,6} & \Leftrightarrow
\begin{aligned}
&X_3 \perp X_5 \mid X_2=2 \\
&X_5 \perp X_3 \mid X_2=2,X_4
\end{aligned}.
\end{aligned}
\end{equation*}
The above statements on the right correspond to setting certain of the MPL-associated conditional distributions identical. This is achieved by partitioning the outcome space of the Markov blankets accordingly. To give an example of how this works in practice, consider the outcome space of the Markov blanket of variable $X_5$ which is made up of $(X_2,X_3,X_4)$. In this case,
\[
X_5 \perp X_2 \mid X_3=0,X_4 \Rightarrow X_5 \perp X_2 \mid X_3=0,X_4=0,
\]
which implies that the $(X_2,X_3,X_4)$-configurations $(0,0,0),(1,0,0)$, and $(2,0,0)$ give rise to identical conditional distributions over $X_5$ and are thereby placed in the same class. Note that restrictions may overlap and care must be taken when constructing the partition. In addition to the above restriction, consider  
\[
X_5 \perp X_3 \mid X_2=1,X_4 \Rightarrow X_5 \perp X_3 \mid X_2=1,X_4=0,
\]
which implies that the $(X_2,X_3,X_4)$-configurations $(1,0,0),(1,1,0)$, and $(1,2,0)$ give rise to identical conditional distributions over $X_5$ and are thereby placed in the same class. However, since both of the considered sets of configurations cover the same configuration, $(1,0,0)$, it implies that all of the mentioned configurations give rise to the same conditional distribution and are thereby placed in the same class. In Figure \ref{fig:oc_part} we show how the final complete partition can be obtained through a stepwise procedure. 

Similar to \eqref{eq:mpl_fac}, the MPL for CMNs factorizes into node-wise marginal conditional likelihoods 
\[
\hat{p}(\mathbf{x}\mid G,\mathcal{C})=\prod_{j=1}^{d} p(\mathbf{x}_j \mid \mathbf{x}_{mb(j)},\scal(j)),
\]
where $\scal(j)=\{ \scal(i,j) \}_{i \in mb(j)}$ denotes the set of edge contexts of edges between node $j$ and its Markov blanket. The set $\scal(j)$ is sufficient for determining how the outcome space of the Markov blanket of node $j$ is partitioned. If all edge contexts associated with a node are empty, the marginal conditional likelihood of the node is reduced to regular MPL calculation where each Markov blanket configuration is considered a separate class.

It has been shown that the MPL criterion is a consistent estimator for the graph structure \citep{PensarMPL}. In the following theorem we establish that the modified MPL criterion is also a consistent estimator of the dependence structure of CMNs.
\begin{theorem}\label{thm:consist_res}
Let $(G^{*},\mathcal{C}^{*},P^{*})$ be a contextual Markov network over d variables for which $(G^{*},\mathcal{C}^{*})$ is maximal regular and $P^{*}$ is faithful to $(G^{*},\mathcal{C}^{*})$. Let $\mathbf{x}$ be a sample of size $n$ generated from $P^{*}$. The local MPL estimator
\begin{equation*}
(\skew{6}\widehat{m}b(j),\skew{3}\widehat{\scal}(j)) = \underset{mb(j),\sscal(j)}{\arg \max} \ p(\mathbf{x}_{j} \mid \mathbf{x}_{mb(j)},\scal(j))
\end{equation*}
is consistent in the sense that $\skew{6}\widehat{m}b(j)=mb^{*}(j)$ and $\skew{3}\widehat{\scal}(j)=\scal^{*}(j)$ eventually almost surely as $n\to\infty$ for $j=1,\ldots,d$. Consequently, the global MPL estimator
\begin{equation*}
(\widehat{G},\widehat{\mathcal{C}}) = \underset{G,\mathcal{C}}{\arg \max} \ \hat{p}(\mathbf{x} \mid G,\mathcal{C})
\end{equation*}
 is consistent in the sense that $\widehat{G}=G^{*}$ and $\widehat{\mathcal{C}}=\mathcal{C}^{*}$ eventually almost surely as $n\to\infty$.
\end{theorem}
\begin{proof}\let\qed\relax
See appendix.
\end{proof}
The consistency property validates the use of MPL from a theoretical perspective, however, it is even more important to study its performance in practice for limited sample sizes. For this purpose, we perform a small-scale simulation study on both synthetic and real-world datasets in Section \ref{sec:num_res}.  

In addition to the MPL part, the final score also contains a structure prior. Previous research has shown that when learning graphical CSI models by optimizing the marginal likelihood (or the BIC), the resulting models tend to be very dense with complex dependence structures \citep{Pensar15,Nyman15b,Pensar15b}. In addition, such models may suffer from poor out-of-sample performance \citep{Pensar15,Pensar15b}. Therefore, a variety of priors have been proposed to further regulate the model fit \citep{Friedman96,Nyman14,Pensar15,Nyman15b,Pensar15b}. Due to the similarity between the marginal likelihood for Bayesian networks with local CSIs and the MPL for CMNs, one would expect the MPL to suffer from the same problem if it is not properly regulated. 

The prior of a CMN structure can be factorized into a prior over the graph and the context given the graph,
\[
p(G,\mathcal{C})=p(\mathcal{C}\mid G) \cdot p(G).
\]
Similar to \citet{Pensar15}, we assume a uniform prior over the graphs and instead design a prior that penalizes inclusion of context elements. More specifically, we construct our prior according to   
\[
p(\mathcal{C}\mid G) \propto \prod_{\{i,j\}\in E} \kappa^{|\sscal(i,j)|(r_i-1)(r_j-1)} = \prod_{j=1}^{d}\prod_{i\in mb(j)} \kappa^{\frac{1}{2}|\sscal(i,j)|(r_i-1)(r_j-1)}
\]
where $|\scal(i,j)|$ is the number of elements in the edge context, $(r_i-1)(r_j-1)$ is a cardinality dependent factor, and $\kappa \in (0,1]$ is a tuning parameter. The  value on $\kappa$ specifies how strongly a context element must be supported by the MPL to be included, $\kappa=1$ results in a non-penalizing uniform prior whereas a small enough $\kappa=\epsilon$ prohibits non-empty edge contexts altogether and thereby reduces the learning procedure to regular MPL learning of the graph structure alone. The problem with a tunable prior is that we must determine a value on $\kappa$ through some form of empirical method. For example, to choose among several candidate values, \citet{Pensar15} used a cross-validation based method. Here we propose a computationally less expensive approach where we simply choose the model with the highest BIC score from a set of models identified under various values on $\kappa$. This is made possible by the fact that we do not use BIC as our objective function during the learning process. The idea is that any potential overfitting with respect to the MPL will be reflected in a reduced BIC score, this would obviously not be the case if the candidate structures were optimized with respect to the BIC. In the experiment section, we study how the chosen models perform in practice both in- and out-of-sample.

\subsection{Search algorithm}
Given a score, we still need to construct a search algorithm for finding high-scoring networks. Since this is an intractable problem, we construct a simple hill-climb method similar to the one used by \citet{PensarMPL}. The idea is to begin from the empty graph and then traverse between neighboring graphs in a greedy manner until a local maximum is reached. The set of neighbors of a graph $G$ is denoted by $\mathcal{N}(G)$ and defined as all graphs that can be reached by adding/deleting a single edge to/from $G$. To include the context structure in the search procedure, we apply a second hill-climb method to identify a context for each considered graph. The empty context is set as the initial state and the search method proceeds by adding elements to the edge contexts in a greedy manner until no further improvement can be achieved by adding a single element. Note that the only elements that need to be evaluated, after the first step, are those of edges overlapping with the edge of the most recently altered context. Furthermore, such elements need only be re-evaluated in terms of the node-wise scores of the nodes part of the edge whose context was changed. Pseudo-code of the search procedures is presented in Algorithm \ref{alg:hc_ug} and \ref{alg:hc_ctxt}. 

\begin{algorithm}\small{
\textbf{Procedure} Graph-Hill-Climb(

\hspace{1cm}$\mathbf{x}$\phantom{$G$} \ \ //\emph{Complete dataset}

\hspace{1cm})

1: \ \ $G,\widehat{G}\leftarrow$ \emph{empty graph}, $\widehat{\mathcal{C}} \leftarrow$ \emph{empty context}\vspace{0.1cm}

2: \ \ \textbf{while} $\widehat{G}$ has changed\vspace{0.1cm}

3: \hspace{0.5cm} $G\leftarrow \widehat{G}$\vspace{0.1cm} 

4: \hspace{0.5cm} \textbf{for each} $G'\in \mathcal{N}(G)$\vspace{0.1cm}  

5: \hspace{1cm} $\mathcal{C}'\leftarrow\text{Context-Hill-Climb}(\mathbf{x},G')$\vspace{0.1cm}  

6: \hspace{1cm} \textbf{if} $\hat p(\mathbf{x},G',\mathcal{C}') > \hat p(\mathbf{x},\widehat{G},\widehat{\mathcal{C}})$\vspace{0.1cm} 

7: \hspace{1.5cm} $(\widehat{G},\widehat{\mathcal{C}}) \leftarrow (G',\mathcal{C}')$\vspace{0.1cm} 

8: \hspace{1cm} \textbf{end}\vspace{0.1cm} 

9: \hspace{0.5cm} \textbf{end}\vspace{0.1cm} 

10:\hspace{0.05cm} \textbf{end}\vspace{0.1cm} 

11:\hspace{0.05cm} \textbf{return} $\widehat{G},\widehat{\mathcal{C}}$\vspace{0.1cm}} 
\caption{Procedure for finding a graph with an associated context structure. \label{alg:hc_ug}}
\end{algorithm}
\begin{algorithm}\small{
\textbf{Procedure} Context-Hill-Climb(

\hspace{1cm}$\mathbf{x}$\phantom{$G$} \ \ //\emph{Complete dataset}

\hspace{1cm}$G$\phantom{$\mathbf{x}$} \ \ //\emph{Graph}

\hspace{1cm})

1: \ \ $\mathcal{C},\widehat{\mathcal{C}}\leftarrow$ \emph{empty context}\vspace{0.1cm} 

2: \ \ \textbf{while} $\widehat{\mathcal{C}}$ has changed\vspace{0.1cm}

3: \hspace{0.5cm} $\mathcal{C}\leftarrow \widehat{\mathcal{C}}$\vspace{0.1cm} 

4: \hspace{0.5cm} \textbf{for each} $(i,j)\in E$\vspace{0.1cm}  

5: \hspace{1cm} \textbf{for each} $x_{cn(i,j)}\in \{\mathcal{X}_{cn(i,j)}\setminus  \scal(i,j)\}$\vspace{0.1cm}  

6: \hspace{1.5cm} $\scal'(i,j)\leftarrow \scal(i,j) \cup x_{cn(i,j)}$\vspace{0.1cm}  

7: \hspace{1.5cm} \textbf{if} $\hat p(\mathbf{x},G,\mathcal{C}') > \hat p(\mathbf{x},G,\widehat{\mathcal{C}})$\vspace{0.1cm} 

8: \hspace{2.5cm} $\widehat{\mathcal{C}} \leftarrow \mathcal{C}'$\vspace{0.1cm} 

9: \hspace{1.5cm} \textbf{end}\vspace{0.1cm} 

10: \hspace{0.9cm} \textbf{end}\vspace{0.1cm} 

11: \hspace{0.4cm} \textbf{end}\vspace{0.1cm} 

12:\hspace{0.05cm} \textbf{end}\vspace{0.1cm} 

13:\hspace{0.05cm} \textbf{return} $\widehat{\mathcal{C}}$\vspace{0.1cm}} 
\caption{Procedure for finding a context structure of a given graph.\label{alg:hc_ctxt}}
\end{algorithm}

\section{Numerical results\label{sec:num_res}}
In this section we perform a simulation study in order to investigate how MPL performs in practice in learning CMN structures for both synthetic and real-world data. In the first part, we look closer into the behavior of the MPL using synthetic datasets generated from a model containing actual CSIs. In the second part, we perform experiments on several real-world datasets in order to investigate if our method is able to find CSIs in real data and thereby improve model quality. To evaluate how well a model fits the training data, we use a standardized BIC score (sBIC) which is divided by the sample size $n$. In addition, we consider the out-of-sample predictive accuracy of the corresponding models. The model parameters are estimated by the MLEs. To enable computation of the MLEs using the conjugate gradient ascent technique described in \citet[][Section A.5.2]{Koller09}, we restricted the experiments to small-scale models.

We ran the CMN search algorithm under different priors specified by
\[
\kappa \in \{\epsilon,n^{-1},n^{-1/2},n^{-1/4}\}.
\]
where $\epsilon$ is a value small enough to prevent edge contexts from being added, this is essentially the same as skipping the context search (step 5 in Algorithm \ref{alg:hc_ug}). We compare our models against traditional non-decomposable Markov networks (MN), which are obtained as part of the CMN search when $\kappa=\epsilon$, as well as decomposable stratified graphical models (SGM), which are learned using the stochastic search algorithm by \citet{Nyman14}, the number of iterations was set to 500 for both the graph and the label procedure.

\subsection{Synthetic data}
In this section we perform simulations on synthetic data generated from a known model that contains CSIs. Having access to the true model allows us to test our method in a controlled setting over a range of sample sizes. Moreover, it allows us to assess the out-of-sample performance using the Kullback-Leibler (KL) divergence from the true distribution $P$ to the approximated distribution $\hat P$ under a model,
\[
D_{KL}(P,\hat P)=\sum_{x\in\mathcal{X}}p(x)\log \frac{p(x)}{\hat p(x)}.
\]
The lower the value, the closer the approximate distribution is to the true distribution. 

As our true model, we use an SGM (originally used in \citet{Nyman14}) which is, as explained earlier, a special case of the CMN model class. The model contains seven binary variables and its dependence structure is illustrated by the labeled graph in Figure \ref{fig:sim_synth_struct}(a). Ten datasets were generated for sample sizes ranging from $250$ to $4000$ (see Supplementary material).

\begin{figure}
\begin{subfigure}{0.24\textwidth}
\begin{center}
\begin{tikzpicture}
\node[at={(0,0)},circle,draw,fill=node_color,line width=0.25mm,minimum size=20pt](6){6};
\node[at={(2,0)},circle,draw,fill=node_color,line width=0.25mm,minimum size=20pt](7){7};
\node[at={(1,1.5)},circle,draw,fill=node_color,line width=0.25mm,minimum size=20pt](5){5};
\node[at={(0,3)},circle,draw,fill=node_color,line width=0.25mm,minimum size=20pt](3){3};
\node[at={(2,3)},circle,draw,fill=node_color,line width=0.25mm,minimum size=20pt](4){4};
\node[at={(0,5)},circle,draw,fill=node_color,line width=0.25mm,minimum size=20pt](1){1};
\node[at={(2,5)},circle,draw,fill=node_color,line width=0.25mm,minimum size=20pt](2){2};
\path (1) edge[line width=0.25mm] node[above=-2] {\small{$10$}} (2);
\path (1) edge[line width=0.25mm] node[left=-2] {\small{$1\hspace{-0.01cm}*$}} (3);
\path (1) edge[line width=0.25mm] (4);
\path (2) edge[line width=0.25mm] (3);
\path (2) edge[line width=0.25mm] (4);
\path (3) edge[line width=0.25mm] (4);
\path (3) edge[line width=0.25mm] (5);
\path (4) edge[line width=0.25mm] node[below right=-2] {\small{$0$}} (5);
\path (5) edge[line width=0.25mm] (6);
\path (5) edge[line width=0.25mm] node[above right=-2] {\small{$0$}} (7);
\path (6) edge[line width=0.25mm] node[below=-1] {\small{$1$}} (7);
\end{tikzpicture}
\end{center}
\caption{True structure.\label{fig:gen_sg}}
\end{subfigure}
\begin{subfigure}{0.24\textwidth}
\begin{center}
\begin{tikzpicture}
\node[at={(0,0)},circle,draw,fill=node_color,line width=0.25mm,minimum size=20pt](6){6};
\node[at={(2,0)},circle,draw,fill=node_color,line width=0.25mm,minimum size=20pt](7){7};
\node[at={(1,1.5)},circle,draw,fill=node_color,line width=0.25mm,minimum size=20pt](5){5};
\node[at={(0,3)},circle,draw,fill=node_color,line width=0.25mm,minimum size=20pt](3){3};
\node[at={(2,3)},circle,draw,fill=node_color,line width=0.25mm,minimum size=20pt](4){4};
\node[at={(0,5)},circle,draw,fill=node_color,line width=0.25mm,minimum size=20pt](1){1};
\node[at={(2,5)},circle,draw,fill=node_color,line width=0.25mm,minimum size=20pt](2){2};
\path (1) edge[line width=0.25mm] (4);
\path (2) edge[line width=0.25mm] (3);
\path (2) edge[line width=0.25mm] (4);
\path (3) edge[line width=0.25mm] node[below=-1] {\small{$*1$}} (4);
\path (3) edge[line width=0.25mm] (5);
\path (4) edge[line width=0.25mm] node[below right=-2] {\small{$0$}} (5);
\path (5) edge[line width=0.25mm] (6);
\path (5) edge[line width=0.25mm] node[above right=-2] {\small{$0$}} (7);
\path (6) edge[line width=0.25mm] node[below=-1] {\small{$1$}} (7);
\end{tikzpicture}
\end{center}
\caption{CMN: $n=250$.\label{fig:synth_n250_cmn}}
\end{subfigure}
\begin{subfigure}{0.24\textwidth}
\begin{center}
\begin{tikzpicture}
\node[at={(0,0)},circle,draw,fill=node_color,line width=0.25mm,minimum size=20pt](6){6};
\node[at={(2,0)},circle,draw,fill=node_color,line width=0.25mm,minimum size=20pt](7){7};
\node[at={(1,1.5)},circle,draw,fill=node_color,line width=0.25mm,minimum size=20pt](5){5};
\node[at={(0,3)},circle,draw,fill=node_color,line width=0.25mm,minimum size=20pt](3){3};
\node[at={(2,3)},circle,draw,fill=node_color,line width=0.25mm,minimum size=20pt](4){4};
\node[at={(0,5)},circle,draw,fill=node_color,line width=0.25mm,minimum size=20pt](1){1};
\node[at={(2,5)},circle,draw,fill=node_color,line width=0.25mm,minimum size=20pt](2){2};
\path (1) edge[line width=0.25mm] node[above=-2] {\small{$10$}} (2);
\path (1) edge[line width=0.25mm] node[left=-2] {\small{$1\hspace{-0.01cm}*$}} (3);
\path (1) edge[line width=0.25mm] (4);
\path (2) edge[line width=0.25mm] (3);
\path (2) edge[line width=0.25mm] (4);
\path (3) edge[line width=0.25mm] node[below=-3] {\small{$\stack{*01}{111}$}} (4);
\path (3) edge[line width=0.25mm] (5);
\path (4) edge[line width=0.25mm] node[below right=-2] {\small{$0$}} (5);
\path (5) edge[line width=0.25mm] (6);
\path (5) edge[line width=0.25mm] node[above right=-2] {\small{$0$}} (7);
\path (6) edge[line width=0.25mm] node[below=-1] {\small{$1$}} (7);
\end{tikzpicture}
\end{center}
\caption{CMN: $n=1000$.\label{fig:synth_n1000_cmn}}
\end{subfigure}
\begin{subfigure}{0.24\textwidth}
\begin{center}
\begin{tikzpicture}
\node[at={(0,0)},circle,draw,fill=node_color,line width=0.25mm,minimum size=20pt](6){6};
\node[at={(2,0)},circle,draw,fill=node_color,line width=0.25mm,minimum size=20pt](7){7};
\node[at={(1,1.5)},circle,draw,fill=node_color,line width=0.25mm,minimum size=20pt](5){5};
\node[at={(0,3)},circle,draw,fill=node_color,line width=0.25mm,minimum size=20pt](3){3};
\node[at={(2,3)},circle,draw,fill=node_color,line width=0.25mm,minimum size=20pt](4){4};
\node[at={(0,5)},circle,draw,fill=node_color,line width=0.25mm,minimum size=20pt](1){1};
\node[at={(2,5)},circle,draw,fill=node_color,line width=0.25mm,minimum size=20pt](2){2};
\path (1) edge[line width=0.25mm] node[above=-2] {\small{$10$}} (2);
\path (1) edge[line width=0.25mm] node[left=-2] {\small{$1\hspace{-0.01cm}*$}} (3);
\path (1) edge[line width=0.25mm] (4);
\path (2) edge[line width=0.25mm] (3);
\path (2) edge[line width=0.25mm] (4);
\path (3) edge[line width=0.25mm] node[below=-2] {$*\hspace{-0.075cm}*\hspace{-0.075cm}1$} (4);
\path (3) edge[line width=0.25mm] (5);
\path (4) edge[line width=0.25mm] node[below right=-2] {\small{$0$}} (5);
\path (5) edge[line width=0.25mm] (6);
\path (5) edge[line width=0.25mm] node[above right=-2] {\small{$0$}} (7);
\path (6) edge[line width=0.25mm] node[below=-1] {\small{$1$}} (7);
\end{tikzpicture}
\end{center}
\caption{CMN: $n=4000$.\label{fig:synth_n4000_cmn}}
\end{subfigure}

\hspace{0.24\textwidth}
\begin{subfigure}{0.24\textwidth}
\vspace{0.5cm}
\begin{center}
\begin{tikzpicture}
\node[at={(0,0)},circle,draw,fill=node_color,line width=0.25mm,minimum size=20pt](6){6};
\node[at={(2,0)},circle,draw,fill=node_color,line width=0.25mm,minimum size=20pt](7){7};
\node[at={(1,1.5)},circle,draw,fill=node_color,line width=0.25mm,minimum size=20pt](5){5};
\node[at={(0,3)},circle,draw,fill=node_color,line width=0.25mm,minimum size=20pt](3){3};
\node[at={(2,3)},circle,draw,fill=node_color,line width=0.25mm,minimum size=20pt](4){4};
\node[at={(0,5)},circle,draw,fill=node_color,line width=0.25mm,minimum size=20pt](1){1};
\node[at={(2,5)},circle,draw,fill=node_color,line width=0.25mm,minimum size=20pt](2){2};
\path (3) edge[line width=0.25mm] (4);
\path (1) edge[line width=0.25mm] node[left=-2] {\small{$1$}} (3);
\path (1) edge[line width=0.25mm] (4);
\path (2) edge[line width=0.25mm] (3);
\path (2) edge[line width=0.25mm] (4);
\path (3) edge[line width=0.25mm] (5);
\path (4) edge[line width=0.25mm] node[below right=-2] {\small{$0$}} (5);
\path (5) edge[line width=0.25mm] (6);
\path (5) edge[line width=0.25mm] node[above right=-2] {\small{$0$}} (7);
\path (6) edge[line width=0.25mm] node[below=-1] {\small{$1$}} (7);
\end{tikzpicture}
\end{center}
\caption{SGM: $n=250$.\label{fig:synth_n250_sgm}}
\end{subfigure}
\begin{subfigure}{0.24\textwidth}
\vspace{0.5cm}
\begin{center}
\begin{tikzpicture}
\node[at={(0,0)},circle,draw,fill=node_color,line width=0.25mm,minimum size=20pt](6){6};
\node[at={(2,0)},circle,draw,fill=node_color,line width=0.25mm,minimum size=20pt](7){7};
\node[at={(1,1.5)},circle,draw,fill=node_color,line width=0.25mm,minimum size=20pt](5){5};
\node[at={(0,3)},circle,draw,fill=node_color,line width=0.25mm,minimum size=20pt](3){3};
\node[at={(2,3)},circle,draw,fill=node_color,line width=0.25mm,minimum size=20pt](4){4};
\node[at={(0,5)},circle,draw,fill=node_color,line width=0.25mm,minimum size=20pt](1){1};
\node[at={(2,5)},circle,draw,fill=node_color,line width=0.25mm,minimum size=20pt](2){2};
\path (1) edge[line width=0.25mm] node[above=-2] {\small{$10$}} (2);
\path (1) edge[line width=0.25mm] node[left=-2] {\small{$1\hspace{-0.01cm}*$}} (3);
\path (1) edge[line width=0.25mm] node[above=7] {\small{$01$}\phantom{00}} (4);
\path (2) edge[line width=0.25mm] (3);
\path (2) edge[line width=0.25mm] (4);
\path (3) edge[line width=0.25mm] (4);
\path (3) edge[line width=0.25mm] (5);
\path (4) edge[line width=0.25mm] node[below right=-2] {\small{$0$}} (5);
\path (5) edge[line width=0.25mm] (6);
\path (5) edge[line width=0.25mm] node[above right=-2] {\small{$0$}} (7);
\path (6) edge[line width=0.25mm] node[below=-1] {\small{$1$}} (7);
\end{tikzpicture}
\end{center}
\caption{SGM: $n=1000$.\label{fig:synth_n1000_sgm}}
\end{subfigure}
\begin{subfigure}{0.24\textwidth}
\vspace{0.5cm}
\begin{center}
\begin{tikzpicture}
\node[at={(0,0)},circle,draw,fill=node_color,line width=0.25mm,minimum size=20pt](6){6};
\node[at={(2,0)},circle,draw,fill=node_color,line width=0.25mm,minimum size=20pt](7){7};
\node[at={(1,1.5)},circle,draw,fill=node_color,line width=0.25mm,minimum size=20pt](5){5};
\node[at={(0,3)},circle,draw,fill=node_color,line width=0.25mm,minimum size=20pt](3){3};
\node[at={(2,3)},circle,draw,fill=node_color,line width=0.25mm,minimum size=20pt](4){4};
\node[at={(0,5)},circle,draw,fill=node_color,line width=0.25mm,minimum size=20pt](1){1};
\node[at={(2,5)},circle,draw,fill=node_color,line width=0.25mm,minimum size=20pt](2){2};
\path (1) edge[line width=0.25mm] node[above=-2] {\small{$10$}} (2);
\path (1) edge[line width=0.25mm] node[left=-2] {\small{$1\hspace{-0.01cm}*$}} (3);
\path (1) edge[line width=0.25mm] (4);
\path (2) edge[line width=0.25mm] (3);
\path (2) edge[line width=0.25mm] (4);
\path (3) edge[line width=0.25mm] (4);
\path (3) edge[line width=0.25mm] (5);
\path (4) edge[line width=0.25mm] node[below right=-2] {\small{$0$}} (5);
\path (5) edge[line width=0.25mm] (6);
\path (5) edge[line width=0.25mm] node[above right=-2] {\small{$0$}} (7);
\path (6) edge[line width=0.25mm] node[below=-1] {\small{$1$}} (7);
\end{tikzpicture}
\end{center}
\caption{SGM: $n=4000$.\label{fig:synth_n4000_sgm}}
\end{subfigure}
\caption{Labeled graphs: (a) true structure, (b)--(d) identified CMN structures for $n=250,1000,4000$, and (e)--(g) identified SGM structures for $n=250,1000,4000$.\label{fig:sim_synth_struct}}
\end{figure}
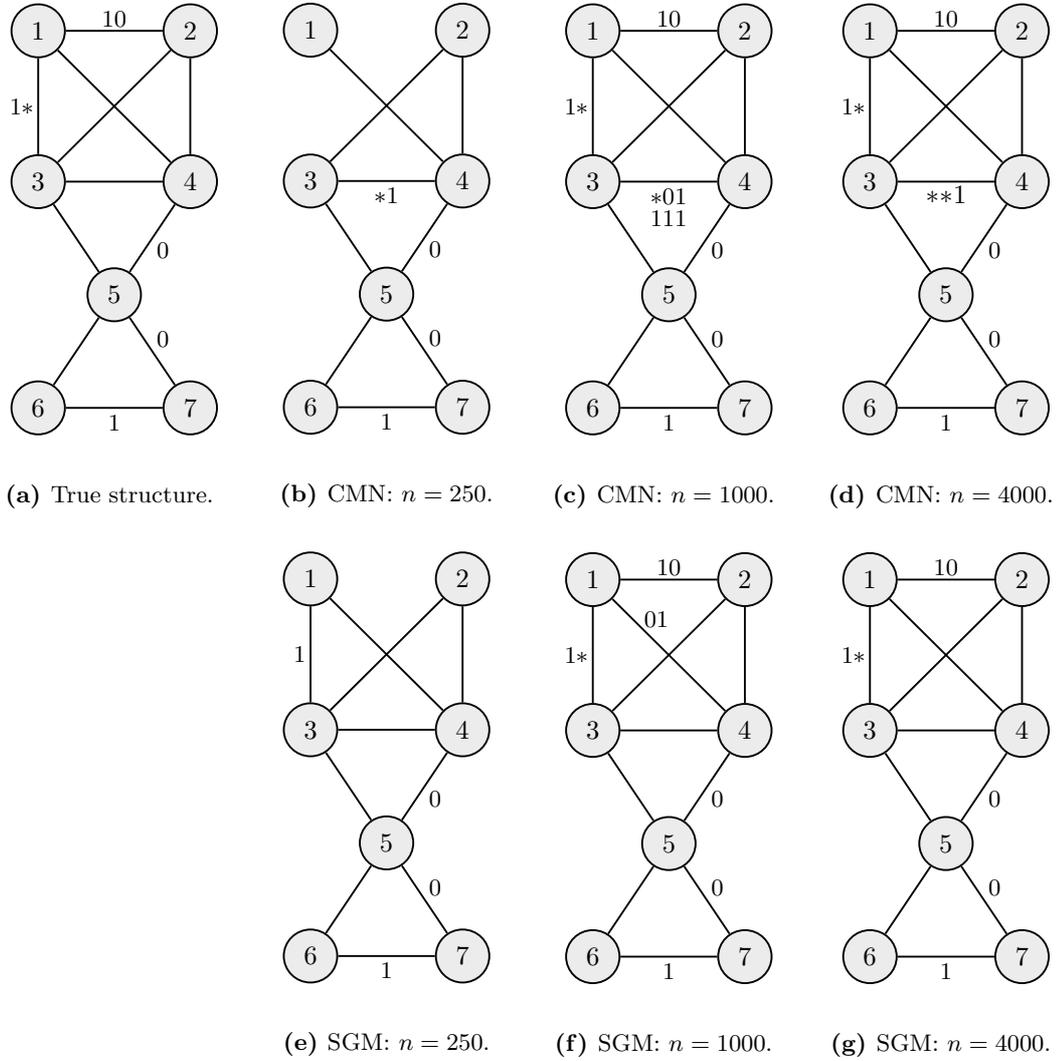
We begin by visually inspecting and comparing the identified model structures, which varied slightly depending on the sample. However, as the sample size was increased, the differences became less distinct. In Figures \ref{fig:sim_synth_struct}(b)--(d) and \ref{fig:sim_synth_struct}(e)--(g), representative structures for various sample sizes for CMNs and SGMs, respectively, are illustrated. It is clear that the resemblance between the identified and correct structure increased with an increased amount of training data for both model classes. For $n=250$, the CMN is missing two edges whereas the SGM is missing only one edge. For $n=1000$, both the CMN and the SGM have the correct underlying graph but they have also an additional label. For $n=4000$, the correct SGM is identified. Interestingly, even for $n=4000$ the CMN still has an additional label at edge $\{ 3,4 \}$. The label represents the edge context $\scal (3,4)=\mathcal{X}_1\times \mathcal{X}_2 \times \{ 1 \}$ which in turn (in combination with the graph) implies that
\[
X_3 \perp X_4 \mid X_1,X_2,X_5=1.
\]
To investigate this further, we need to look closer at the underlying distribution. Therefore, in Tables \ref{tab:ex_cpds}(a)--(b) we have listed the conditional distributions of $X_3$ and $X_4$ given their Markov blanket for all configurations where $X_5=1$. The tables show that the conditional distributions of $X_3$ and $X_4$ are very similar regardless of the value on $X_4$ and $X_3$, respectively, when $X_5=1$. Hence, the additional label can be expected to enjoy a rather high support due to the lack of dependence even though it is not included in the true model structure, in fact, it is not even a valid label in an SGM. Finally, note also the truly identical conditional distributions on row 3 and 7 as well as 4 and 8 in Table \ref{tab:ex_cpds}(a) which are due to the context of edge $\{ 1,3 \}$.
\begin{table}
\caption{Conditional probability distributions of (a) $X_3$ and (b) $X_4$ given their Markov blankets and the context $X_5=1$.\label{tab:ex_cpds}}
\begin{subtable}{.5\textwidth}
\begin{center}
\small{
\begin{tabular}{c@{\hskip 0.2cm} c@{\hskip 0.2cm} c@{\hskip 0.2cm} c c}
\toprule
$X_1$ & $X_2$ & $X_4$ & $X_5$ & $p(X_3\mid X_{1,2,4,5})$ \\
\midrule
0 & 0 & 0 & 1 & 0.8140 \ \ 0.1860 \\
0 & 0 & 1 & 1 & 0.7887 \ \ 0.2113 \\
0 & 1 & 0 & 1 & 0.3333 \ \ 0.6667 \\
0 & 1 & 1 & 1 & 0.3478 \ \ 0.6522\\
1 & 0 & 0 & 1 & 0.9459 \ \ 0.0541 \\
1 & 0 & 1 & 1 & 0.9289 \ \ 0.0711 \\
1 & 1 & 0 & 1 & 0.3333 \ \ 0.6667 \\
1 & 1 & 1 & 1 & 0.3478 \ \ 0.6522 \\
\bottomrule
\end{tabular}
}
\end{center}
\caption{\label{tab:ex_cpd1}}
\end{subtable}
\begin{subtable}{.5\textwidth}
\begin{center}
\small{
\begin{tabular}{c@{\hskip 0.2cm} c@{\hskip 0.2cm} c@{\hskip 0.2cm} c c}
\toprule
$X_1$ & $X_2$ & $X_3$ & $X_5$ & $p(X_4\mid X_{1,2,3,5})$ \\
\midrule
0 & 0 & 0 & 1 & 0.7143 \ \ 0.2857 \\
0 & 0 & 1 & 1 & 0.6809 \ \ 0.3191 \\
0 & 1 & 0 & 1 & 0.5000 \ \ 0.5000 \\
0 & 1 & 1 & 1 & 0.5161 \ \ 0.4839 \\
1 & 0 & 0 & 1 & 0.5172 \ \ 0.4828 \\
1 & 0 & 1 & 1 & 0.4444 \ \ 0.5556 \\
1 & 1 & 0 & 1 & 0.0588 \ \ 0.9412 \\
1 & 1 & 1 & 1 & 0.0625 \ \ 0.9375 \\
\bottomrule
\end{tabular}
}
\end{center}
\caption{ }
\end{subtable}
\end{table}

Next we examine the predictive properties of the identified models. In Tables \ref{tab:synth_res}(a)--(b) we have listed the average sBIC scores and KL divergences for the considered model classes. Throughout the range of sample sizes, CMN obtained the highest sBIC scores, SGM came in second and MN third. This indicates that our method is able to identify sound CSIs and learn CMN structures that fit the training data better than not only traditional MNs, but also decomposable SGMs learned using the marginal likelihood. In terms of KL divergence, the CMNs obtained lower values than the MNs for all sample sizes and lower values than the SGMs for the two largest sample sizes. Consequently, the improved model fit did not translate into an improved out-of-sample performance for the three smallest sample sizes when comparing with SGMs. However, this is perhaps not too surprising considering that the underlying model is an SGM, next we move on to real-world datasets. 

\begin{table}
\caption{Summary of the results obtained in the synthetic data experiments. The bold font indicates the highest value of each row in terms of sBIC and the lowest value in terms of KL divergence.\label{tab:synth_res}}
\begin{center}
\small{
\begin{tabular}{c@{\hskip 0.8cm} c@{\hskip 0.3cm} c@{\hskip 0.3cm} c c@{\hskip 0.6cm} c@{\hskip 0.3cm} c@{\hskip 0.3cm} c@{\hskip 0.3cm}}
\toprule
\multirow{2}{*}{$n$} & \multicolumn{3}{c}{sBIC} & & \multicolumn{3}{c}{KL divergence} \\
& CMN & MN & SGM & & CMN & MN & SGM \\
\midrule
$250$ & \textbf{-4.3696} & -4.4095 & -4.3755  & & 0.0867 & 0.0922 & \textbf{0.0658}  \\
$500$ & \textbf{-4.3274} & -4.3604 & -4.3327 & & 0.0435 & 0.0472 & \textbf{0.0273}  \\
$1000$ & \textbf{-4.3069} & -4.3302 & -4.3119 & & 0.0142 & 0.0201 & \textbf{0.0129}  \\
$2000$ & \textbf{-4.2789} & -4.2930 & -4.2825  & & \textbf{0.0052} & 0.0070 & 0.0055 \\
$4000$ & \textbf{-4.2667} & -4.2740 & -4.2689  & & \textbf{0.0026} & 0.0032 & 0.0027 \\
\bottomrule
\end{tabular}
}
\end{center}
\end{table}

\subsection{Real-world data}
In this section we investigate how our method performs on a collection of real-world datasets. Since we now do not have access to a true model, we assess the out-of-sample performance by calculating the predictive probability of a hold-out test set. More specifically, we split each dataset $\mathbf{x}$ into ten subsets $\mathbf{x}_{(1)},\ldots,\mathbf{x}_{(10)}$ of similar size and in turn use each subset as test data and the remaining sets as training data. The model structure and parameters are learned from the training data and the standardized log-probability of the test data under the identified model is calculated. The final predictive accuracy measure is given by the average over the ten partitions.

For our experiments, we have chosen ten datasets that have been used extensively in previous research. The datasets, which are listed in Table \ref{tab:real_data_descr} in the appendix, have been pre-processed in order to fit our criteria of being complete, discrete, and small-scale. Five of the datasets are binary and five are non-binary. We have omitted the SGM results for the non-binary datasets since the current implementation only allows for binary variables. For more information about the datasets and the pre-processing, see Table \ref{tab:real_data_descr}.

The results of the experiments are summarized in Table \ref{tab:real_res}. For all of the considered datasets, CMN obtained higher sBIC scores than its competitors. This confirms our observations from the previous section. The improved model fit now also resulted in improved out-of-sample predictive accuracy, CMN obtained a higher predictive accuracy than SGM for all the binary datasets. Compared with MN, CMN obtained a higher predictive accuracy for eight out of the ten datasets. In terms of the structural properties listed in Table \ref{tab:real_res2}, the identified CMNs  contained more edges than the MNs but fewer than the SGMs. The CMNs required the smallest number of parameters.

As a result of how the final model is chosen, our method will prefer models that fit the training data well. Still, to evaluate how well the BIC measure performs in choosing models with high predictive accuracy, in Table \ref{tab:real_kappa} we compare our CMN results against results obtained when keeping the $\kappa$-value fixed. Overall, the BIC method outperformed the fixed-value priors illustrating the advantage of our case-wise approach as opposed to specifying a single prior. The last column in Table \ref{tab:real_kappa}, which corresponds to models learned under the weakest prior, shows tendencies of overfitting with respect to the MPL for some of the datasets. Although the models are equally good as the BIC optimal models for half of the datasets, the overall performance over all the datasets is poor.

\begin{table}
\caption{Summary of the results obtained in the real-world data experiments. The five datasets above the dotted line are binary and the five datasets below are non-binary. The bold font indicates the highest value of each row for the property in question.\label{tab:real_res}}
\begin{center}
\small{
\begin{tabular}{l@{\hskip 0.6cm} c@{\hskip 0.3cm} c@{\hskip 0.3cm} c c@{\hskip 0.6cm} c@{\hskip 0.3cm} c@{\hskip 0.3cm} c@{\hskip 0.3cm}}
\toprule
\multirow{2}{*}{Name} & \multicolumn{3}{c}{sBIC} & & \multicolumn{3}{c}{Predictive accuracy} \\
& CMN & MN & SGM & & CMN & MN & SGM \\
\midrule
Congressional voting & \textbf{-6.4800} & -6.5068 & -6.6583  & & -6.4526 & \textbf{-6.4151} & -6.5428  \\
Coronary heart disease & \textbf{-3.6506} & -3.6567 & -3.6522  & & \textbf{-3.6395} & -3.6460 & -3.6417  \\
Economic activity & \textbf{-4.0077} & -4.0413  & -4.0513 & & \textbf{-3.9515} & -3.9698 & -4.0245  \\
Finnish parliament & \textbf{-7.3767} & -7.4031 & -7.4124  & & \textbf{-7.3525} & -7.3579 & -7.3623  \\
Women in mathematics\vspace{0.1cm} & \textbf{-3.7770} & -3.7829 & -3.7771  & & \textbf{-3.7547} & -3.7561 & -3.7564  \\
\hdashline
\vspace{-0.25cm}\\
Car evaluation & \textbf{-7.9225} & -7.9691 & - & & \textbf{-7.7651} & -7.7690 & -  \\
Contraceptive method & \textbf{-8.3405} & -8.3568 & - & & \textbf{-8.2108} & -8.2188 & -  \\
Mushroom & \textbf{-5.0416} & -5.1035 & -  & & -4.4276 & \textbf{-4.4262} & - \\
Soybean & \textbf{-9.3404} & -9.3773 & -  & & \textbf{-9.0416} & -9.1259 & - \\
Wisconsin breast cancer & \textbf{-5.0012} & -5.1344 & - & & \textbf{-4.9057} & -5.0388 & -  \\
\bottomrule
\end{tabular}
}
\end{center}
\end{table}
\begin{table}
\caption{Number of edges and parameters of the models in the real-world data experiments. The five datasets above the dotted line are binary and the five datasets below are non-binary. \label{tab:real_res2}}
\begin{center}
\small{
\begin{tabular}{l@{\hskip 0.6cm} c@{\hskip 0.3cm} c@{\hskip 0.3cm} c c@{\hskip 0.6cm} c@{\hskip 0.3cm} c@{\hskip 0.3cm} c}
\toprule
\multirow{2}{*}{Name} & \multicolumn{3}{c}{Edges} & & \multicolumn{3}{c}{Parameters} \\
& CMN & MN & SGM & & CMN & MN & SGM \\
\midrule
Congressional voting & 24.9 & 23.4 & 30.0 & & 40.0 & 40.0 & 54.3  \\
Coronary heart disease & 5.5 & 4.5 & 6.5 & & 9.5 & 10.9 & 12.1  \\
Economic activity & 14.3 & 11.6  & 12.8 & & 18.1 & 22.2 & 23.2  \\
Finnish parliament & 31.9 & 26.2 & 34.8 & & 42.4 & 48.0 & 61.6  \\
Women in mathematics\vspace{0.1cm} & 6.0 & 4.9 & 6.4 & & 11.1 & 11.8 & 11.2  \\
\hdashline
\vspace{-0.25cm}\\
Car evaluation & 6.0 & 6.0 & - &  & 79.0 & 100.0 & -  \\
Contraceptive method & 10.9 & 10.5 & - & & 83.2 & 84.7 & - \\
Mushroom & 17.0 & 17.7 & - & & 756.0 & 832.0 & -  \\
Soybean & 14.1 & 12.4 & - & & 58.9 & 60.8 & -  \\
Wisconsin breast cancer & 13.3 & 12 & - &  & 57.3 & 62.6 & -  \\
\bottomrule
\end{tabular}
}
\end{center}
\end{table}
\begin{table}
\caption{Predictive accuracy of models identified under different priors. The column $\kappa_{\text{\tiny{BIC}}}$ represents our current approach where the BIC optimal model is chosen. The bold font indicates the highest value of each row. \label{tab:real_kappa}}
\begin{center}
\small{
\begin{tabular}{l c@{\hskip 0.6cm} c@{\hskip 0.6cm} c@{\hskip 0.4cm} c@{\hskip 0.3cm} c }
\toprule
Name & $\kappa_{\text{\tiny{BIC}}}$ & $\kappa=\epsilon$ & $\kappa=n^{-1}$ & $\kappa=n^{-1/2}$ & $\kappa=n^{-1/4} $  \\
\midrule
Car evaluation & \textbf{-7.7651} & -7.7690 & -7.7690 & -7.7700 & \textbf{-7.7651}   \\
Congressional voting & -6.4526 & \textbf{-6.4151} & \textbf{-6.4151} & -6.4900 & -6.9397  \\
Contraceptive method & \textbf{-8.2108} & -8.2188 & -8.2188 & -8.2188 & -8.2249  \\
Coronary heart disease & \textbf{-3.6395} & -3.6460 & -3.6460 & -3.6437 & \textbf{-3.6395}  \\
Economic activity & \textbf{-3.9515} & -3.9698 & -3.9698 & -3.9648 & \textbf{-3.9515}  \\
Finnish parliament & \textbf{-7.3525} & -7.3579 & -7.3593 & -7.3580 & -7.4324 \\
Mushroom & -4.4276 & -4.4262 & -4.4262 & \textbf{-4.4259} & -4.4276  \\
Soybean & \textbf{-9.0416} & -9.1259 & -9.1259 & -9.1524 & -9.1168  \\
Wisconsin breast cancer & \textbf{-4.9057} & -5.0388  & -5.0388 & -5.0079 & -4.9187 \\
Women in mathematics\vspace{0.1cm} & \textbf{-3.7547} & -3.7561 & -3.7563 & -3.7551 & \textbf{-3.7547}  \\
\hdashline
\vspace{-0.25cm}\\
Average & \textbf{-5.9502} & -5.9724 & -5.9725 & -5.9787 & -6.0171 \\
\bottomrule
\end{tabular}
}
\end{center}
\end{table}

\section{Conclusions}

In this work we have extended the class of traditional Markov networks (MNs) by introducing contextual Markov networks (CMNs). In addition to the graph, the dependence structure of a CMN is specified by a collection of edge contexts. An edge context specifies cases in which the direct influence of the edge is cut off according to the notion of context-specific independence (CSI). In line with previous work, we showed that the additional CSI constraints can be accounted for through linear restrictions on the log-linear parameters.

The class of CMNs is a slightly extended version of previous model classes \citep{Corander03,Nyman14,Nyman15b}. One of the main challenges in earlier research has been learning the dependence structure of the models in an efficient manner without imposing  artificial restrictions, such as chordality. The main contribution of this work is extending the scope of marginal pseudo-likelihood (MPL) to also cover CMNs. This was achieved by combining the concept of marginal pseudo-likelihood for Markov networks \citep{PensarMPL} with the concept of local CSIs in Bayesian network learning \citep{Boutilier96,Friedman96,Pensar15}. The resulting objective function has a tractable closed-form expression and the corresponding estimator was here proven to be consistent in the large sample limit.

The results in our experiments showed that MPL is an excellent candidate criterion for learning CMN structures. The obtained CMNs enjoyed an improved predictive accuracy both in- and out-of-sample when compared with ordinary MNs. Moreover, CMN also outperformed the previously introduced class of decomposable stratified graphical models (SGMs) on several real-world datasets. Due to computational reasons concerning parameter estimation in non-decomposable models, we restricted the experiments to small-scale models. Still, the potential of our scoring method makes it an attractive candidate for considering CMNs in high-dimensional real-world applications, for example, in modeling the dependence structure among features in CMN-based classifiers \citep{Nyman15a}. 

Having a closed-form objective function enables learning of non-chordal CMNs for large-scale systems. A natural next step in future research will thus be to scale up the dimension of the models. First of all, this will require a more efficient search algorithm. In this work we preferred simplicity over efficiency, since the main goal was to prove the concept of MPL as an objective function for CMN structure learning. To improve the efficiency of future algorithms, one could further exploit the MPL factorization. For example, rather than re-doing the context search for all edges after each edge change, one could limit the search to edges for which the common neighbors have been modified. If necessary, one could also perform a pre-scan to identify eligible edges and thereby reduce the space of candidate graphs \citep{PensarMPL}.

Another current limitation in CMN learning is the task of estimating the model parameters. Whereas parameter estimation poses no particular problem for decomposable models, in a CMN it quickly becomes infeasible as the dimension of the model is increased. A possible solution would be to replace the likelihood with a more tractable objective function such as the pseudo-likelihood \citep{Besag75}. Another attractive approach, which allows for truly high-dimensional parameter estimation, would be to learn the parameters in a distributed manner where the global problem is divided into smaller independent sub-problems. The sub-problems can then be solved in parallel and the final solution is constructed from the solutions of the separate sub-problems \citep{Liu12,Mizrahi14}. Future research will show which of these alternatives provides the best scalability without imposing excessive demands on the implementation.

\section*{Acknowledgements}
JP was supported by the Magnus Ehrnrooth Foundation and the Finnish Doctoral Programme in Stochastics and Statistics. HN was supported by the Foundation of Åbo Akademi University, as part of the grant for the Center of Excellence in Optimization and Systems Engineering. JC was supported by the Academy of Finland grant 251170.

\section*{Supporting information} Additional material for this article is available online including a MATLAB package containing functions for learning the structure and parameters of a CMN.

\bibliographystyle{ScandJbst}
\bibliography{MPL_cmn}
\vspace{0.5cm}
\noindent Johan Pensar, Department of Mathematics and Statistics, Åbo Akademi University, Finland. \\
E-mail: johan.pensar@abo.fi

\section*{Appendix}
\begin{proof}\textbf{(Proposition \ref{thm:cn})}
Let $X_{V}$ be a set of variables satisfying the Markov properties of the undirected graph $G=(V,E)$. Furthermore, let $i,j,k\in V$ such that $\{i,j\}\in E$ and assume, for example, that $\{ i,k \}\not\in E$ (since $ k\not\in cn(i,j)$). The pairwise Markov property states that
\begin{equation}\label{eq:ci}
X_{i} \perp X_{k} \mid X_{V\setminus \{ i,k \} },
\end{equation}
which is equivalent to the conditions
\[
p(X_{i} \mid x_{k},X_{V\setminus \{ i,k \}})=p(X_{i} \mid x'_{k},X_{V\setminus \{ i,k \}})\text{ for all } x_k,x'_k\in \mathcal{X}_k.
\]
Now consider the CSI statement 
\begin{equation}\label{eq:csi}
X_{i} \perp X_{j} \mid x_{cn(i,j)},x_{k},X_{V\setminus \{ \overbar{cn}(i,j)\cup k\} },
\end{equation}
which is equivalent to the conditions
\[
p(X_{i} \mid x_{j},x_{cn(i,j)},x_{k},X_{V\setminus \{ \overbar{cn}(i,j)\cup k\}})=p(X_{i} \mid x'_{j},x_{cn(i,j)},x_{k},X_{V\setminus \{ \overbar{cn}(i,j)\cup k\}})\text{ for all } x_j,x'_j\in \mathcal{X}_j.
\]
By combining \eqref{eq:ci} and \eqref{eq:csi} we obtain
\begin{equation*}\label{eq:g_eq}
\begin{aligned}
p(X_{i}\mid x_{j},x_{cn(i,j)},x'_{k},X_{V\setminus \{ \overbar{cn}(i,j)\cup k\}})&\overset{\eqref{eq:ci}}{=} p(X_{i}\mid x_{j},x_{cn(i,j)},x_{k},X_{V\setminus \{ \overbar{cn}(i,j)\cup k\}})\\
&\overset{\eqref{eq:csi}}{=} p(X_{i}\mid x'_{j},x_{cn(i,j)},x_{k},X_{V\setminus \{ \overbar{cn}(i,j)\cup k\}})\\
&\overset{\eqref{eq:ci}}{=} p(X_{i}\mid x'_{j},x_{cn(i,j)},x'_{k},X_{V\setminus \{ \overbar{cn}(i,j)\cup k\}}),
\end{aligned}
\end{equation*}
which holds for all $x_j,x'_j\in \mathcal{X}_j$ and for all $x'_k\in\mathcal{X}_k$. Consequently, we can rewrite the above conditions as 
\[
p(X_{i} \mid x_{j},x_{cn(i,j)},X_{V\setminus  \overbar{cn}(i,j) })=p(X_{i} \mid x'_{j},x_{cn(i,j)},X_{V\setminus  \overbar{cn}(i,j)})\text{ for all } x_j,x'_j\in \mathcal{X}_j,
\]
which is equivalent to the CSI statement
\begin{equation*}\label{eq:csi2}
X_{i} \perp X_{j} \mid x_{cn(i,j)},X_{V\setminus \overbar{cn}(i,j)}.
\end{equation*}
Consequently, under the Markov properties of the graph, the CSI in equation \eqref{eq:csi} induce the same restrictions on the dependence structure over the variables as the above CSI. Note that we would reach the same conclusion by assuming $\{ j,k \}\not\in E$. Note also that the same reasoning as above holds for $cn(i,j)=\varnothing$, however, in this case the considered edge would be removed as a result of the induced restrictions which correspond to a conditional independence statement \citep[see][Figure 3]{Nyman15b}.
\end{proof}

\begin{proof}\textbf{(Proposition \ref{thm:ec_loglin_rest})}
The proof follows the same outline as the work of \citet{Nyman15b}, however, here we consider a general setting with non-binary variables and non-chordal graphs in combination with the modified definition of the considered CSIs. An element in an edge context, $x_{cn(i,j)}\in \scal(i,j)$, implies that for all $x_{V\setminus \overbar{cn}(i,j)}\in \mathcal{X}_{V\setminus \overbar{cn}(i,j)}$ and $x_j,x'_{j}\in\mathcal{X}_{j}$, it holds that
\begin{equation}\label{eq:edge_ctxt_rest}
p(X_{i} \mid x_{j},x_{cn(i,j)},x_{V\setminus \overbar{cn}(i,j)})=p(X_{i} \mid x'_{j},x_{cn(i,j)},x_{V\setminus \overbar{cn}(i,j)})\\
\end{equation}
which is equivalent to the conditions
\[
\frac{p(x_{i}\mid x_{j},x_{cn(i,j)},x_{V\setminus \overbar{cn}(i,j)})}{p(x'_{i}\mid x_{j},x_{cn(i,j)},x_{V\setminus \overbar{cn}(i,j)})}=\frac{p(x_{i}\mid x'_{j},x_{cn(i,j)},x_{V\setminus \overbar{cn}(i,j)})}{p(x'_{i}\mid x'_{j},x_{cn(i,j)},x_{V\setminus \overbar{cn}(i,j)})} \text{ for all } x_i,x'_i \in \mathcal{X}_i.
\]
By noting that
\[
\frac{p(x_{i}\mid x_{j},x_{cn(i,j)},x_{V\setminus \overbar{cn}(i,j)})}{p(x'_{i}\mid x_{j},x_{cn(i,j)},x_{V\setminus \overbar{cn}(i,j)})}=\frac{p(x_{i}, x_{j},x_{cn(i,j)},x_{V\setminus \overbar{cn}(i,j)})}{p(x'_{i}, x_{j},x_{cn(i,j)},x_{V\setminus \overbar{cn}(i,j)})}
\] 
and taking the logarithm of both sides, we can rewrite the initial condition \eqref{eq:edge_ctxt_rest} in terms of the log-linear parameterization such that
\[
\sum_{A\subseteq cn(i,j)}\hspace{-0.2cm}\left(\phi_{A\cup \{ i,j \}}(x_{A},x_{i},x_{j})-\phi_{A\cup \{ i,j \}}(x_{A},x'_{i},x_{j})\right)=\hspace{-0.3cm}\sum_{A\subseteq cn(i,j)} \hspace{-0.2cm} \left(\phi_{A\cup \{ i,j \}}(x_{A},x_{i},x'_{j})-\phi_{A\cup \{ i,j \}}(x_{A},x'_{i},x'_{j})\right)
\]
for all $x_{j},x'_{j}\in\mathcal{X}_{j}$ and $x_{i},x'_{i}\in\mathcal{X}_i$. For the particular case $(x_i,x_j)=(0,0)$, all of the above $\phi$-functions except terms containing $(x'_i,x'_j)$ will be equal to zero due to restriction \eqref{eq:zero_rest}. This gives us the reduced set of restrictions
\[
\sum_{A\subseteq cn(i,j)}\phi_{A\cup \{ i,j \}}(x_{A},x'_{i},x'_{j})=0 \text{ for all } x'_{\{i,j\}}\in \{1,\ldots,r_{i}-1\}\times\{1,\ldots,r_{j}-1\}
\]
where $\phi_{A\cup \{ i,j \}}(x_{A},x'_{i},x'_{j})=0$ if $x_{k}=0$ for any $k\in A$. A closer examination of the reduced set of restrictions reveals that it is equivalent to the original set of restrictions and is thereby sufficient for capturing \eqref{eq:edge_ctxt_rest}.
\end{proof}

\begin{proof}\textbf{(Theorem \ref{thm:consist_res})}
\citet{PensarMPL} showed that MPL is consistent in identifying the correct graph for traditional MNs. Considering maximal regular structures, the MPL is also consistent in identifying the correct graph of a CMN. This conclusion is based on the fact that the influence of an edge in a maximal regular structure cannot completely vanish. Thereby, the maximum conditional likelihood of a node will eventually be achieved given all of the true Markov blanket members. Moreover, adding false members will not improve the conditional likelihood in the large sample limit since a CMN follows the Markov properties of its graph. It thereby remains to show that MPL is consistent in identifying the correct edge contexts. This is done by following a similar reasoning as above, however, we now provide more details.

We proceed by investigating the asymptotic behavior of the node-wise log-score
\begin{equation*}
\log p(\mathbf{x}_{j} \mid \mathbf{x}_{mb(j)},\scal(j))=\sum_{l=1}^{q_j}\left(\log\Gamma(\alpha_{jl})-\log\Gamma(n_{jl}+\alpha_{jl}) +\sum_{i=1}^{r_j}\left( \log\Gamma(n_{ijl}+\alpha_{ijl})-\log\Gamma(\alpha_{ijl})\right)\right)
\end{equation*}
where the $l$-index indicates the $q_j$ distinct classes specified by $\scal (j)$. By using Stirling's asymptotic formula and following the same approach as in \citet{PensarMPL}, we obtain the asymptotic result
\begin{equation*}
\log p(\mathbf{x}_{j} \mid \mathbf{x}_{mb(j)},\scal(j)) \to \sum_{l=1}^{q_{j}}\sum_{i=1}^{r_{j}} n_{ijl} \log \frac{ n_{ijl}}{ n_{jl}}- \frac{(r_{j}-1)q_{j}}{2}\log n +O(1).
\end{equation*}
as $n\to \infty$. Note that $\{n_{ijl}/n_{jl}\}_{i=1}^{r_j}$ is the MLE of the probabilities specifying the conditional distribution $p(X_j\mid x_{mb(j)}^{(l)})$. Using the above result, we now show that the local MPL estimator of node $j$ will avoid \emph{under-} and \emph{overestimation} of the context structure in the large sample limit, that is, it will not leave out true elements from or add false elements to edge contexts in $\scal (j)$.

\emph{Underestimation:} Consider a situation where an element is missing from an edge context, $x'_{cn(i,j)}\not \in \scal (i,j)$, although it is included in the correct structure, $x'_{cn(i,j)} \in \scal^{*} (i,j)$. For simplicity we assume that the resulting structure is maximal since the induced dependence structure is otherwise identical to the one with the element included. In practice there are situations where it be would necessary to remove additional elements to reach a distinct maximal structure, however, the proof would follow the exact same outline. Let $l_1$ and $l_2$ represent distinct classes in the candidate network which induce identical conditional distributions of variable $j$ and would be merged if adding $x'_{cn(i,j)}$. As $n\to \infty$,
\begin{equation*}
\frac{ n_{ijl_1}}{ n_{jl_1}}-\frac{n_{ijl_2}}{ n_{jl_2}} \to 0
\end{equation*}
for $i=1,\ldots,r_j$. This gives the asymptotic log-factor
\[
\log \frac{p(\mathbf{x}_{j} \mid \mathbf{x}_{mb(j)},\scal(j))}{p(\mathbf{x}_{j} \mid \mathbf{x}_{mb(j)},\scal^{*}(j))} \to (q^{*}_j-q_j)\frac{r_j-1}{2}\log n+O(1)
\]
where $q^{*}_j$ and $q_j$ are the number of parent classes for the respective case. Since $q^{*}_j<q_j$, the log-factor will tend to $-\infty$ confirming that the correct edge context is preferred as $n \to \infty$.

\emph{Overestimation:} Consider a situation where an element is added to an edge context, $x'_{cn(i,j)} \in \scal (i,j)$, although it is not included in the correct structure, $x'_{cn(i,j)} \not\in \scal^{*} (i,j)$. Let $l_1$ and $l_2$ represent distinct classes in the true network which have been merged to a single class in the candidate network by the context element $x'_{cn(i,j)}$. As $n\to \infty$

\begin{equation}\label{eq:lof_fac}
\log \frac{p(\mathbf{x}_{j} \mid \mathbf{x}_{mb(j)},\scal(j))}{p(\mathbf{x}_{j} \mid \mathbf{x}_{mb(j)},\scal^{*}(j))} \to n\cdot \Delta_1+\log n \cdot \Delta_2+O(1)
\end{equation}
where
\[
\Delta_1=\sum_{i=1}^{r_j} \frac{n_{ijl_1}+n_{ijl_2}}{n} \log\frac{n_{ijl_1}+n_{ijl_2}}{n_{jl_1}+n_{jl_2}}-\sum_{i=1}^{r_j}\left( \frac{n_{ijl_1}}{n} \log\frac{n_{ijl_1}}{n_{jl_1}}+\frac{n_{ijl_2}}{n}\log\frac{n_{ijl_2}}{n_{jl_2}}\right)
\]
tends to a constant, which is a guaranteed to be negative by the log sum inequality since
\begin{equation*}
\frac{ n_{ijl_1}}{ n_{jl_1}}-\frac{n_{ijl_2}}{ n_{jl_2}} \not\to 0
\end{equation*}
for some $i=1,\ldots,r_j$. On the other hand, 
\[
\Delta_2=(q^{*}_j-q_j)\frac{r_j-1}{2}
\]
is a positive constant since $q^{*}_j>q_j$. However, since the first term in \eqref{eq:lof_fac} grows at a faster rate than the second, the log-factor will tend to $-\infty$ confirming that the correct edge context is preferred as $n \to \infty$.

The above results concerning under- and overestimation together confirm that the MPL will eventually, as the sample size is increased, support merging of truly identical conditional distributions while keeping non-identical distributions separate. In other words, the correct edge contexts associated with a node will eventually be identified as optimal by the local MPL estimator as $n\to\infty$. Consequently, the global MPL estimator is consistent in identifying the correct CMN structure. 
\end{proof}

\begin{table}[H]
\caption{Description and references for the real-world datasets used in this work. All observations with missing values were removed. Any additional pre-processing  is listed in the last column as comments.\label{tab:real_data_descr}}
\begin{center}
\footnotesize{
\begin{tabular}{@{\hskip 0.05cm} l@{\hskip 0.2cm} c@{\hskip 0.2cm} c@{\hskip 0.2cm} c@{\hskip 0.2cm} l@{\hskip 0.2cm} l@{\hskip 0.05cm}}
\toprule
Name & $d$ & $n$ & $|\mathcal{X}_d|$ & For more information, see... & Comments \\
\midrule
Car evaluation & 7 & 1728 & 6912 & \citet{Lichman13}& -  \\
Congressional voting & 15 & 312 & 32768 & \citet{Lichman13} & Variables 3 and 7 were removed. \\
Contraceptive method & 10 & 1473 & 55296 & \citet{Lichman13}& Variable 1 categorization: \\
& & & & & $(\leq31)\rightarrow 0, (32-37)\rightarrow 1, (\geq 38)\rightarrow 2$ \\
& & & & & Variable 4 categorization: \\
& & & & & $(0)\rightarrow 0, (1-2)\rightarrow 1, (\geq 3)\rightarrow 2$ \\
Coronary heart disease & 6 & 1841 & 64  & \citet{Edwards85} & - \\
Economic activity & 8 & 665 & 256 & \citet{Whittaker90} & - \\
Finnish parliament & 15 & 1806 & 32768 & \citet{Nyman14} & Variables 1--15 were selected from\\
& & & & & the original dataset.\\
Mushroom & 8 & 5644 & 21504 & \citet{Lichman13} & Variables 1--8 were selected from \\
& & & & &the original dataset.\\
Soybean & 11 & 266 & 62208 & \citet{Lichman13} & Variables 3--13 were selected from \\
& & & & &the original dataset.\\
Wisconsin breast cancer & 10 & 683 & 17496 & \citet{Lichman13} & Variables 1--9 categorization: \\
& & & & &  $(\leq3)\rightarrow 0, (4-7)\rightarrow 1, (\geq 8)\rightarrow 2$ \\
Women in mathematics & 6 & 1190 & 64  & \citet{Lacampagne79} & - \\
\bottomrule
\end{tabular}
}
\end{center}
\end{table}

\end{document}